\documentclass[a4paper,fleqn,usenatbib]{mnras}

\usepackage{newtxtext,newtxmath}

\usepackage[T1]{fontenc}
\usepackage{ae,aecompl}
\usepackage{mathtools}
\usepackage{subfigure}
\usepackage[labelfont=bf]{caption}

\usepackage{graphicx}	
\usepackage{amsmath}	
\usepackage{amssymb}	

\newcommand{\beq}{\begin{equation}}
\newcommand{\eeq}{\end{equation}}
\def\la{\buildrel < \over {_{\sim}}}
\def\ga{\buildrel > \over {_{\sim}}}

\title[Particle escape in SNRs]{Exploring particle escape in supernova remnants through gamma rays}
\author[Celli, Morlino, Gabici \& Aharonian]{
	S.~Celli, $^{1,2,3}$  \thanks{silvia.celli@roma1.infn.it}
	G.~Morlino,$^{4,2}$ 
	S.~Gabici$^{5}$
	and F.~A.~Aharonian$^{2,3,6}$
\\
	$^{1}$Dipartimento di Fisica dell'Universit\`a La Sapienza, P.le Aldo Moro 2, 00185 Roma, Italy\\
	$^{2}$Gran Sasso Science Institute, Viale Francesco Crispi 7, 67100 L'Aquila, Italy \\
	$^{3}$Max-Planck-Institut f\"ur Kernphysik, Postfach 103980, D-69029 Heidelberg, Germany\\
	$^{4}$INAF, Osservatorio Astrofisico di Arcetri, L.go E.~Fermi 5, 50125 Firenze, Italy  \\
	$^{5}$APC, AstroParticule et Cosmologie, Universit\'e Paris Diderot, CNRS/IN2P3, CEA/Irfu, Obs de Paris, Sorbonne Paris Cit\'e, France \\
	$^{6}$Dublin Institute for Advanced Studies, 31 Fitzwilliam Place, Dublin 2, Ireland  \\
}

\begin{document}

\date{Accepted -----. Received -----}


\maketitle

\label{firstpage}

\begin{abstract}
The escape process of particles accelerated at supernova remnant (SNR) shocks is one of the poorly understood aspects of the shock acceleration theory. Here we adopt a phenomenological approach to study the particle escape and its impact on the gamma-ray spectrum resulting from hadronic collisions both inside and outside of a middle-aged SNR.
Under the assumption that in the spatial region immediately outside of the remnant the diffusion coefficient is suppressed with respect to the average Galactic one, we show that a significant fraction of particles are still located inside the SNR long time after their nominal release from the acceleration region. This fact results into a gamma-ray spectrum that resembles a broken power law, similar to those observed in several middle-aged SNRs. Above the break, the spectral steepening is determined by the diffusion coefficient outside of the SNR and by the time dependence of maximum energy.
Consequently, the comparison between the model prediction and actual data will contribute to determining these two quantities, the former being particularly relevant within the predictions of the gamma-ray emission from the halo of escaping particles around SNRs which could be detected with future Cherenkov telescope facilities.
We also calculate the spectrum of run-away particles injected into the Galaxy by an individual remnant. Assuming that the acceleration stops before the SNR enters the snowplow phase, we show that the released spectrum can be a featureless power law only if the accelerated spectrum is $\propto p^{-\alpha}$ with $\alpha>4$.
\end{abstract}


\section{Introduction}  
\label{sec:intro}

Understanding the escape of accelerated particles from expanding spherical shocks is a key ingredient to establishing a connection between SNRs and the origin of Galactic  cosmic rays (CRs). It is often assumed that the spectrum of particles released into the Galaxy by a single SNR resembles the instantaneous spectrum of particles accelerated at the shock. According to the predictions of diffusive shock acceleration theory (DSA), such a spectrum is a featureless power law in energy $E^{-\alpha}$ with slope $\alpha \approx 2$ over a very broad energy interval \cite[see, e.g., reviews by][]{malkovDrury2001, Blasi(rev)2013}. The validity of such assumption depends on several subtleties of the acceleration process, i.e. {\it i}) the amount of time that particles spend inside the SNR, during which they would suffer severe adiabatic losses, {\it ii}) the rate at which particles of different energy are released from the SNR at each time, and {\it iii}) the temporal evolution of the acceleration efficiency during the remnant evolution. 

In a scenario where particles are confined inside the remnant until it dissolves into the interstellar medium (ISM), these would lose a substantial fraction of their energy because of the adiabatic expansion of the shocked plasma. 
On the other hand, the observation of the knee in the CR spectrum at a particle energy of few PeV suggests that the sources of Galactic CRs should be able to inject in the ISM particles up to at least such energies.
This implies that, in order to compensate for adiabatic energy losses, SNR shocks should in fact be able to accelerate particles well beyond the PeV domain, which seems so prohibitive \citep{lagage} that this scenario does not appear to be realistic.

A more realistic, though still qualitative picture for the particle escape emerges from the fact that SNR shocks slow down as the mass of the ISM swept up by the shock increases. During the Sedov-Taylor (adiabatic or ST) phase \citep{taylor1950,sedovBook}, the shock radius expands with time as $t^{0.4}$, which is slower than the $t^{0.5}$ root mean square displacement of CRs expected if their transport is governed by spatial diffusion. In such a scenario, particles start to diffuse away from the shock and the probability that they might return to it from upstream is gradually reduced \cite[see, e.g.][]{Drury2011}. The dilution of particles over large volumes also reduces their capability of exciting magnetic turbulence upstream of the shock due to various plasma instabilities. Less turbulence means less confinement of particles at shocks, and therefore at some point CRs will become completely decoupled from the shock and will escape the SNR. Even though there is a broad consensus on the fact that the escape should be energy dependent -- higher energy particles escaping the shock earlier -- the details of such a process are still not well understood.

In fact, a similar reasoning may be applied also to describe particle escape during the ejecta-dominated phase, which precedes the Sedov one and is characterised by a very mild deceleration of the SNR shock \citep{Chevalier:1982, Truelove-McKee:1999}. While in this case the expansion rate of the SNR shell is larger than the spatial diffusion rate of CRs (the shock radius scales as $t^s$ with $s > 0.5$), particles of very high energy might still escape the SNR. This is because even a mild deceleration of the shock suffices to reduce appreciably the effectiveness of CR streaming instability. 
In addition, also the non-resonant instability, often invoked as the main mechanism for the magnetic field amplification in the ejecta dominated phase, requires that a sizeable fraction of particles at the highest energy should escape in order for the instability to be effective \citep{Bell2004, Bell+2013, Bell&Schure2014, Amato-Blasi:2009}.
Nevertheless, the behavior of particle escape during this phase is not particularly relevant to the objectives of this paper, as a very minor fraction of particles is expected to escape the shock in this way. 

After the decoupling from the SNR, the transport of particles will be determined by the properties of the ambient magnetic turbulence. In fact, the same particles that are escaping from the SNR could generate the magnetic turbulence by means of plasma instabilities such as the streaming instability \citep{Skilling1971} as well as the non-resonant instability \citep{Bell2004}.
In such a scenario, the diffusion coefficient outside of the remnant $D_{\rm out}$ might be suppressed in the transition region between the shock  and the unperturbed ISM with respect to the average Galactic coefficient $D_{\rm Gal} (p)\simeq 10^{28} (pc/10~\textrm{GeV})^{1/3}$~cm$^2$~s$^{-1}$ \cite[e.g.][]{maurin2014}. It is hence clear that one of the main uncertainty of modeling the escape process concerns the value of $D_{\rm out}$. In principle, there is no reason why $D_{\rm out}$ should be equal to $D_{\rm Gal}$: in fact, the latter is usually inferred from secondary over primary CR ratios and it represents an average value over the particle propagation time within the whole magnetic halo of the Galaxy, hence it could be very different from the diffusion coefficient inside the Galactic Plane. Given that a theoretical determination of $D_{\rm out}$ is challenging, one may wonder whether it could be possible to constrain this physical quantity by means of observations, particularly in the high-energy (HE) and very-high-energy (VHE) gamma-ray domain, and to provide some understanding concerning how the escape mechanism works by means of a phenomenological approach towards the existing gamma-ray measurements of SNRs. Indeed, gamma-ray emission is expected from the vicinity of SNRs due to the interactions of escaping particles with ambient gas, especially (but not only) if the gas is structured in massive molecular clouds \citep{Gabici+2009}. The study of such emission is of paramount importance because it allows us to directly observe a manifestation of particle escape from shocks, and to constrain this poorly understood aspect of particle acceleration at SNRs. 

Particle escape from SNR shocks has been the subject of several works, exploring either the connection between run-away particles from the SNR shock and the CR spectrum observed at Earth \citep{Ptuskin-Zirakashvili:2003,Bell+2013,Malkov+2013,Cardillo+2015} or studying the signatures of escaping particles in terms of gamma-ray emission from nearby molecular clouds \citep{Gabici+2009,ohira+2010}. Note that a proper treatment of the escape process is extremely relevant in the search for PeV particle accelerators \citep{GabiciAharonian2007}, as observationally high-energy particles are more likely to be found outside of the SNR shock than inside or in its shell \citep{Aharonian2013}. On the other hand, we are mostly interested in the very initial stages of the escape, when the run-away particles are still located in the close vicinity of the shock, in order to explore the escape conditions through the gamma-ray spectrum detected from the SNR. Given the large uncertainties of current theoretical models aimed at describing particle escape from shocks \citep{Malkov+2013,Nava+2016,DAngelo+2018}, here we will adopt a phenomenological approach. The transport of particles that decoupled from the SNR shock will be described by means of a diffusion coefficient which is both isotropic and spatially homogeneous. Though deviations from this simplest scenario, such as anisotropies and/or spatial variations in the transport of particles, may play an important role \citep{Giacinti+2013,Nava-Gabici2013,DAngelo+2018}, we will neglect them in this work as we aim at describing the radiative signatures from SNRs produced by escaping particle in the most simple scenario. The present study will be limited to middle-aged SNRs, namely remnants evolving through the adiabatic phase, mainly because of two reasons: {\it i}) the amount of escaping particles should be large enough to produce more evident observational effects in secondary gamma rays, and {\it ii}) the remnant hydrodynamical evolution can be well approximated by the ST solution, which allows to provide a simple analytical model for the description of particle propagation. Therefore, the treatment presented does not apply to young SNRs that are still evolving in the ejecta-dominated phase. 

Within a simplified description of the particle transport in spherical symmetry, we obtain a time-dependent analytical solution for the density distribution of both the particles confined by the shock, undergoing acceleration and adiabatic losses, as well as the escaping particles still diffusing in the remnant region. Note that, in order to derive an analytical solution to the particle transport equation, we assume a homogeneous diffusion coefficient and neglect non-linear effects. The obtained solutions depend on the SNR temporal evolution and on the diffusive regime operating at the time when the particles start to escape the shock.  It is therefore possible to quantify the density of particles located within the shock radius and outside of it. Consequently, we derive both the morphology and the spectral energy density of the secondary gamma rays produced at the interaction between the accelerated protons and the target gas, in order to explore the possibility of constraining the regime of operation of particle escape by means of HE and VHE observations. Moreover, as the same escaping particles will eventually contribute to the Galactic cosmic-ray flux, we quantify the flux of run-away particles from middle-aged SNRs, investigating several different acceleration spectra.

The paper is structured as follows. In \S~\ref{sec:model} a simplified model that describes the particle propagation within and around a middle-aged SNR is presented. 
Though the model is not intended to provide a complete description of the particle escape mechanism, it predicts interesting features on the particle spectrum, which are discussed in \S~\ref{sec:protons}. The predicted fluxes of secondary gamma rays produced in hadronic collisions between run-away particles and ambient gas are presented in \S~\ref{sec:gamma}, where the presence of extended TeV halos around SNRs is discussed and their detectability by future generation instruments such as CTA is investigated. In addition, since the escape process is a key ingredient to understand the formation of the Galactic CR flux detected at Earth, the contribution of the run-away particle flux from middle-aged SNRs to the flux of Galactic CRs is evaluated and discussed in \S~\ref{sec:finj}. The results obtained are summarized in \S~\ref{sec:conc}, where conclusions are also derived. 

\section{A simplified model for particle propagation}    
\label{sec:model}

In this Section, we model the propagation of accelerated particles inside and outside a middle-aged SNR in order to properly calculate the spectrum of protons contained in these regions. For the sake of simplicity, we assume spherical symmetry both inside and outside the remnant. We note that the assumption of spherical symmetry outside the SNR is justified in case of highly turbulent medium, like core collapse supernovae (CC-SNe) which expand in the wind-blown bubble produced by their progenitor \citep{Zirakashvili-Ptuskin2018} or in superbubbles, like e.g. the complex Cygnus region \citep{Parizot+2004}. Indeed, simulations of stellar wind-blown bubbles show that the variation of the wind properties during the stellar evolution causes the termination shock to be non stationary and to inject vorticity in the shocked wind \citep{dwarkadas2008}. On the other hand, if a regular magnetic field is present, a cylindrical symmetry would be more suitable (for this scenario the reader is referred to \citet{Nava-Gabici2013,DAngelo+2018}). The transport equation in spherical coordinates describing the evolution of the phase space density $f(t,r,p)$ of accelerated protons reads as
\begin{equation} 
\label{eq:transport}
 \frac{\partial{f}}{\partial{t}} + u \frac{\partial{f}}{\partial{r}} = 
 \frac{1}{r^2} \frac{\partial}{\partial r} \left[ r^2 D \frac{\partial{f}}{\partial{r}}\right]
 + \frac{1}{r^2} \frac{\partial (r^2 u)}{\partial r}  \frac{p}{3} \frac{\partial f}{\partial p} \, ,
\end{equation}
where $u(t,r)$ is the advection velocity of the plasma and $D(t,r,p)$ is the effective spatial diffusion coefficient experienced by particles. In the following, we will solve Eq.~(\ref{eq:transport}) by adopting two different approximations, tailored at describing the propagation of respectively {\it i}) the particles confined inside the remnant, tightly attached to the expanding plasma, and {\it ii}) the non-confined particles, which freely diffuse in the space after having escaped the shock region. In order to solve analytically Eq.~\eqref{eq:transport}, several assumptions will be introduced, concerning: {\it i}) the evolutionary stage of the remnant, as described in \S~\ref{sec:SNR_evolution}, {\it ii}) the particle spectrum accelerated at the shock, which is discussed in \S~\ref{sec:f_shock}, and {\it iii}) the temporal evolution of the particle maximum momentum produced at the shock, that is  explored in \S~\ref{sec:pmax}. Consequently, the results derived in \S~\ref{sec:f_conf} and \S~\ref{sec:f_esc} apply within the range of validity of the aforementioned assumptions.

\subsection{Evolution of middle-aged SNRs}    
\label{sec:SNR_evolution}
We define middle-aged SNRs as those evolving in the ST phase, when the shock slows down as the swept-up matter becomes larger than the mass of the ejecta $M_\textrm{ej}$ while radiative losses are still not significant. Their characteristic age is $T_{\rm SNR} \gtrsim10^4$~yr). During this evolutionary stage, the shock position $R_{\rm sh}$ and the shock speed $u_{\rm sh}$ evolve in time according to the adiabatic solution \citep{sedovBook,Truelove-McKee:1999,matzner1999}, that in the case of a shock expanding through a uniform medium with density $\rho_0$ reads as
\begin{eqnarray} 
 R_{\rm sh}(t) = \left( \xi_0 \frac{E_{\rm SN}}{\rho_0} \right)^{1/5} t^{2/5} \,,			\label{eq:ST_R}	  \\
 u_{\rm sh}(t) = \frac{2}{5} \left( \xi_0 \frac{E_{\rm SN}}{\rho_0} \right)^{1/5} t^{-3/5}	\label{eq:ST_u} \, ,
\end{eqnarray}
where $\xi_0= 2.026$ and $E_{\rm SN}$ represents the kinetic energy released at the supernova (SN) explosion. The time that marks the transition between the ejecta-dominated phase and the ST phase is the so-called Sedov time, namely
\begin{equation} 
   t_\textrm{Sed} \simeq 1.6 \times 10^3 \, \textrm{yr} \left(\frac{E_\textrm{SN}}{10^{51} \, \textrm{erg}} \right)^{-1/2} 
   		\left(\frac{M_\textrm{ej}}{10 \, M\odot} \right)^{5/6} 
		\left(\frac{\rho_0}{1 \, m_\textrm{p}/\textrm{cm}^3} \right)^{-1/3}
\end{equation}
where $m_\textrm{p}$ is the proton mass. The internal structure of the SNR is determined by the hydrodynamical evolution of the moving plasma: in the following, we will adopt the linear velocity approximation introduced by \citet{Ostriker-McKee1988}, in which the plasma velocity profile for $r \leq R_{\rm sh}$ is given by 
\begin{equation} 
\label{eq:u(r,t)}
 u(t,r) =  \left(1-\frac{1}{\sigma} \right) \frac{u_{\rm sh}(t)}{R_{\rm sh}(t)} r \, ,
 \end{equation}
$\sigma$ being the compression ratio at the shock ($\sigma=4$ for strong shocks). 

\subsection{CR distribution at the shock}   
 \label{sec:f_shock}
Following \cite{Ptuskin+2005}, we assume that the efficiency in converting the shock bulk kinetic energy into relativistic particles, $\xi_{\rm CR}$, is constant in time.  
The distribution function of CR accelerated at the shock is determined by DSA and it is predicted to be a featureless power law in momentum with slope $\alpha$. A maximum value of the particle momentum $p_{\max}$, though not naturally embedded in the DSA theory, has to exist in order to limit the spectral energy density of accelerated particles. Such a value is either connected with the accelerator age, that implies a finite time for acceleration, or with the particle escape from the system. In a simplified form, we can write the particle spectrum at the shock as
\begin{equation} 
\label{eq:f_0}
 f_0(t,p) = \frac{3 \, \xi_{\rm CR} u^2_{\rm sh}(t) \rho_0}{4 \pi \, c (m_p c)^4  \Lambda(p_{\max,0}(t))} 
 		\left( \frac{p}{m_\textrm{p} c} \right)^{-\alpha}  \Theta \left[ p_{\max,0}(t)-p \right] \, ,
\end{equation}
where $c$ is the speed of light. We leave the slope $\alpha$ as a free parameter of the model. It is worth to recall, however, that DSA predicts $\alpha$ to be equal or very close to 4. The function $p_{\max,0}(t)$ represents the maximum momentum accelerated at the shock at the time $t$, as will be discussed in the next Section, while $\Lambda(p_{\max,0})$ is required to normalize the spectrum such that the CR pressure at the shock is $P_{\rm CR} = \xi_{\rm CR} \rho_0 u_{\rm sh}^2$. We thus have
\begin{equation} 
\label{eq:Lambda}
  \Lambda(p) = \int_{p_{\min}/m_p c}^{p/m_p c} y^{4-\alpha} \left(1 + y^2\right)^{-1/2} dy \,.
\end{equation}
The fact that the efficiency $\xi_{\rm CR}$ is constant in time is a key element of the whole problem, including the calculation of the final CR spectrum injected by SNRs into the ISM (see \S~\ref{sec:finj}). Such assumption is usually connected with the idea that the acceleration efficiency should saturate at roughly the same level, regardless of the shock speed, provided that the shock is strong. Though a proof of this conjecture is still missing, hints in this direction are provided by particle-in-cell simulations \citep{Caprioli-Spitkovsky2014}. In addition, analytical models which implement the thermal leakage recipe and account for non-linear effects have shown that the efficiency remains constant as long as the condition $M \gg 1$ is fulfilled (see e.g. Fig.~1 in \citet{caprioli2012}).

\subsection{Maximum energy at the shock}	
\label{sec:pmax}
A self-consistent description of the maximum energy achievable in the acceleration mechanism in a non-stationary framework requires the correct modeling of the evolution of the magnetic turbulence, which is supposed to be self-generated by the same accelerated particles and possibly damped through frictional effects and wave cascade. Since such a complete description does not exist yet, we will here use a quite general recipe, often adopted in the literature, which assumes that the maximum momentum increases with time during the free expansion phase, when the shock is actively accelerating particles, and then it decreases during the Sedov-Taylor phase according to a power law in time \cite[see e.g. ][]{Gabici+2009}, namely
\begin{equation} 
\label{eq:pmax0}
 p_{\max,0}(t) =
  \begin{cases} 
   p_\textrm{M} \left( t/t_{\rm Sed} \right)     & \text{if } t \leqslant  t_{\rm Sed} 	\\
   p_\textrm{M} \left( t/t_{\rm Sed} \right)^{-\delta}     & \text{if } t > t_{\rm Sed} \,,
  \end{cases}
\end{equation}
where $p_\textrm{M}$ represents the absolute maximum momentum, achieved at $t=t_{\rm Sed}$. The reason for considering a transition for $p_{\rm max}$ at the Sedov time is connected with the fact that during the free-expansion phase the particles achieve a maximum momentum generally higher than during the adiabatic phase. However, as at this stage the number of accelerated particles is rather low, the average spectrum of escaping particles will result steeper than during the adiabatic phase \citep{Ptuskin+2005}. Hence, the fact that more particles are accelerated during the Sedov stage (the remnant is more extended) has the net effect of producing a peak in $p_{\rm max}$ right at the Sedov time. Note that the evolution of $p_{\rm max}(t)$ during the ED phase is largely uncertain: for this reason, we also explored the case of $p_{\rm max}=p_{\rm M}$ for $t \leqslant  t_{\rm Sed}$, and observed that it produces marginal differences in the gamma-ray spectrum of middle-aged SNRs. In Eq.~\eqref{eq:pmax0}, $\delta$ is a free parameter of the model, bounded to be positive. The value of this parameter strongly depends on the temporal evolution of the magnetic turbulence. In the simple stationary test-particle approach $\delta= 1/5$. This value should be regarded purely as a lower limit, since in a more realistic scenario the strength of the magnetic turbulence is expected to be proportional to some power of the shock speed, which decreases in time (see Appendix \ref{sec:appA} for more details).

By inverting Eq.~(\ref{eq:pmax0}), we can also define the escape time for particles of given momentum $p$, corresponding to the time when these particles cannot be confined anymore by the turbulence and start escaping from the shock. The particle escape time reads as
\begin{equation} 
\label{eq:tesc}
 t_{\rm esc}(p) = t_{\rm Sed} \left( p/p_{\rm M} \right)^{-1/\delta} \, .
\end{equation}
It is also useful to define the escape radius as 
\begin{equation} 
\label{eq:Resc}
 R_{\rm esc}(p) = R_{\rm sh}\left(t_{\rm esc}(p) \right) \, .
\end{equation}
The onset of the escape process in the acceleration scenario introduces a unique feature in the evolution of the particle distribution, that will behave differently before and after $t_{\rm esc}(p)$. In fact, at times smaller than $t_{\rm esc}(p)$, particles closely follow the shock evolution as they are strictly tightened to the turbulence. On the other hand, at later times, when the turbulence starts to fade out, particles behave disconnected by the shock. Particles evolving in these two regimes will be named respectively \emph{confined particles} and \emph{non-confined particles}, as described in \S~\ref{sec:f_conf} and \S~\ref{sec:f_esc}. Note that, while confined particles are only located inside the remnant radius (by definition), the non-confined population can be located both inside and outside the radius, depending on the diffusion conditions operating there. In fact, even if non-confined particles have nominally escaped the shock, they can possibly be scattered towards the remnant interior, and reside there for some time after the escape time, thus producing observable effects in the secondary radiation emitted at hadronic interactions. Later, once the turbulence has reduced significantly, these particles are able to leave the source region, propagate through the ISM and eventually reach the Earth, contributing to the diffuse flux of CRs.

Though Eq.~\eqref{eq:pmax0} may appear too simplistic, it allows to explore the escape mechanism independently on the microphysics of the process. However, we will also explore a situation where a more refined calculation of $p_{\max,0}$ is adopted. In particular we will use the description derived by \citet{schure2013,Bell&Schure2014} and also adopted in \citet{Cardillo+2015}, who considered the possibility that the escaping CRs excite plasma instabilities, leading to the growth of both resonant and non-resonant modes, thus achieving efficient magnetic field amplification and particle scattering. Both the instability channels are driven by the fact that CRs stream at super-alfvenic speed, thus inducing a reaction in the background plasma to restore a null net current. The essential difference between the resonant and non-resonant linear instability is that non-resonant modes result from a collective effect of CRs, namely from their strong drift, while individual CRs are responsible for resonant modes. Considering the non-resonant instability developed by the CR streaming from a remnant expanding into a homogeneous medium, in the assumption that a constant fraction of the shock kinetic energy is instantaneously transferred to the escaping particle flux, one derives the following implicit equation in the maximum energy $E_\textrm{max,0}(t)$

\begin{equation}
\label{eq:emaxCardillo4}
 E_\textrm{max,0}(t) \ln \left( \frac{E_{\max,0}(t)}{E_{\min}}  \right) 
  = \frac{e \sqrt{4\pi \rho_0}}{10c} \xi_\textrm{CR}  u^2_{\rm sh}(t) R_{\rm sh}(t) \, ,
\end{equation}
where $E_\textrm{min}$ is the minimum energy produced by acceleration during the Sedov phase, which does not depend on time, and $e$ is the electron charge. Note that the maximum energy is connected to the maximum momentum by the relation $E_\textrm{max,0}(t)=\sqrt{p^2_\textrm{max,0}(t) c^2+m_\textrm{p}^2 c^4}$. Eq.~\eqref{eq:emaxCardillo4} holds whenever the differential energy spectrum produced during the acceleration is $\propto E^{-2}$, since it was derived by combining Eq.~(2) and Eq.~(9) of \citet{Cardillo+2015} (and setting $m=0$, corresponding to expansion into a homogeneous medium). The approach defined by Eq.~\eqref{eq:emaxCardillo4} implies that the maximum momentum produced at the shock varies with time according to the remnant evolutionary stage: as already discussed in \S~\ref{sec:SNR_evolution}, in the following we will only consider remnants evolving through the ST stage. Correspondingly, within this scenario, the escape time of particles with energy $E$ would be dictated by 
\begin{equation}
\label{eq:tescCardillo4}
  t_\textrm{esc}(E) = \left[ \frac{4\sqrt{\pi \rho_0} e}{125c} \xi_{\rm CR} 
  	\left( \frac{\xi_0 E_{\rm SN}}{\rho_0}  \right)^{3/5} \frac{1}{E \ln(E/E_{\min})}  \right]^{5/4} \, .
\end{equation}
On the other hand, in the case of an acceleration spectrum $\propto E^{-(2+\beta)}$ (with $\beta \neq 0$), the equation regulating $E_\textrm{max,0}(t)$ reads as

\begin{equation}
\label{eq:emaxCardillo43}
E_\textrm{max,0}(t) \left[ \left( \frac{E_\textrm{max,0}(t)}{E_\textrm{min}} \right)^\beta -1 \right] =  \left( \frac{\beta}{1+\beta} \right) \frac{e \sqrt{4\pi \rho_0}}{10c} \xi_\textrm{CR} u^2_{\rm sh}(t) R_{\rm sh}(t) \, ,
\end{equation}
while the escape time is
\begin{equation}
\label{eq:tescCardillo43}
t_\textrm{esc}(E)  = \left[ \frac{4\sqrt{\pi \rho_0}e}{125c} \xi_{\rm CR} \left( \frac{\xi_0 E_{\rm SN}}{\rho_0} \right)^{3/5}  \left( \frac{\beta}{1+\beta} \right) \frac{1}{E}
\left( \frac{E^\beta_\textrm{min}}{E^\beta-E^\beta_\textrm{min}} \right) \right]^{5/4}.
\end{equation}
Note that Eqs.~\eqref{eq:emaxCardillo4} and \eqref{eq:emaxCardillo43} show explicitly the fact that the maximum energy depends on the acceleration efficiency, since the higher is the efficiency, the larger is the current of escaping particles. In addition, these are implicit equations for $E_\textrm{max,0}(t)$, which can be solved with standard numerical techniques.

\subsection{Distribution of confined particles}   
 \label{sec:f_conf}
When $t < t_{\rm esc}(p)$ particles with momentum $p$ are confined inside the SNR and do not escape the shock, due to the wall of turbulence generated at the shock itself. A reasonable approximation for the distribution of these confined particles, that we call $f_{\rm conf}(t,r,p)$ from here on, can be obtained by solving Eq.~(\ref{eq:transport}) in the approximation that the diffusion term can be neglected \citep{Ptuskin+2005}. This is a good approximation if the typical diffusion length is much smaller than the SNR size, namely if $\sqrt{D_\textrm{in} t} \ll R_\textrm{sh}(t)$, which in the ST phase translates into the following condition on the diffusion coefficient inside of the shock
\begin{equation}
D_\textrm{in} \ll 10^{28} \left(\frac{E_{\rm SN}}{10^{51} \rm erg}\right)^{\frac{1}{2}}
						     \left(\frac{M_{\odot}}{M_{\rm ej}} \right)^{\frac{1}{6}} 
						     \left(\frac{n_0}{\rm cm^{-3}} \right)^{-\frac{1}{3}}
						    \left(\frac{t}{t_\textrm{Sed}} \right)^{-\frac{1}{5}} \, {\rm cm^2} {\rm s}^{-1} \, .
\end{equation}
The above condition depends only weakly on $t$ and it is satisfied if $D_{\rm in}(p)$ is suppressed with respect to the average Galactic diffusion coefficient by at least a factor $\sim (p/10 \, {\rm GeV} c^{-1})^{-1/3}$. 
The simplified transport equation for confined particles reads as
\begin{equation} 
\label{eq:transport_in}
 \frac{\partial{f_{\rm conf}}}{\partial{t}} + u \frac{\partial{f_{\rm conf}}}{\partial{r}} = 
 \frac{1}{r^2} \frac{\partial (r^2 u)}{\partial r}  \frac{p}{3} \frac{\partial f_{\rm conf}}{\partial p}
\end{equation}
and its solution can be easily obtained by using the method of characteristics, where the plasma speed inside the SNR is approximated by Eq.~(\ref{eq:u(r,t)}). The solution can be written in the terms of the acceleration spectrum $f_0$ (see Eq.~\eqref{eq:f_0}) as follows:
\begin{equation} 
\label{eq:f_c}
 f_{\rm conf}(t,r,p) = f_0 \left( \left( \frac{R_{\rm sh}(t)}{R_{\rm sh}(t')} \right)^{1-\frac{1}{\sigma}} p, t'(t,r) \right) 
\end{equation}
\cite[see ][]{Ptuskin+2005}, where $t'(t,r)$ represents the time when the plasma layer, that at the time $t$ is located at the position $r$, has been shocked. Such quantity can be obtained from the equation of motion of a plasma layer, i.e. $dr/dt = u(t,r)$, where the velocity profile is given by Eq.~(\ref{eq:u(r,t)}): integrating this equation by parts from $t'$ to $t$, and using Eq.~(\ref{eq:ST_R}), one obtains
\begin{equation} 
\label{eq:t_sh}
 t'(t,r) = \left( \frac{\rho_0}{ \xi_0 E_\textrm{SN}} \right)^{\sigma/2} \, r^{5\sigma/2} \, t^{1-\sigma}  \,.
\end{equation}
We can recast Eq.~(\ref{eq:f_c}) in a simpler form, by using Eqs.~(\ref{eq:ST_R}), (\ref{eq:ST_u}) and (\ref{eq:f_0}) and neglecting the mild dependence of $\Lambda(p_{\max})$ on $t$, thus getting
\begin{equation} 
\label{eq:f_in(t,r)}
 f_{\rm conf}(t,r,p) = f_0(p,t) \left( \frac{t'}{t} \right)^{\epsilon}  \, \frac{\Lambda(t)}{\Lambda(t')} \,
 			       \Theta\left[ p_{\max}(t,r) - p \right] \, ,
\end{equation}
where $\Lambda(t)$ is a shortcut for $\Lambda \left(p_{\max,0}(t) \right)$, while the exponent $\epsilon$ is defined as
\beq
  \epsilon = {\frac{2\alpha (\sigma -1)}{5\sigma} -\frac{6}{5}}  \,.
\eeq
The function $p_{\max}(t,r)$ is the maximum momentum of particles located at position $r$ and time $t$, and it is equal to the maximum momentum of particles accelerated at time $t'$ diminished by adiabatic losses occurred between $t^\prime$ and $t$, i.e.
\begin{equation} 
\label{eq:pmax}
 p_{\max}(t,r) = p_{\max,0}(t') \left( \frac{R_{\rm sh}(t')}{R_{\rm sh}(t)} \right)^{1-\frac{1}{\sigma}} 
 	             =  p_{\max,0}(t) \left( \frac{t'}{t} \right)^{\frac{2(\sigma-1)}{5\sigma} - \delta}  \, ,
\end{equation}
where the last step has been obtained by using Eq.~(\ref{eq:pmax0}) for $t>t_{\rm Sed}$.
The latter equation implies that, for $\delta < \delta^* \equiv 2(\sigma-1)/(5\sigma)$ (namely $\delta^* = 3/10$ for strong shocks), the decrease of the maximum energy at the shock is slower than the decrease of the maximum energy in the remnant interior, as due to adiabatic losses. In such a case, at any given time $t$, particles with momentum $p_{\max,0}(t)$ are only located close to the shock. On the contrary, for $\delta > \delta^*$, at every position $r$ the distribution function is $f(t,r,p_{\max,0}(t)) > 0$. In other words, the condition $\delta > \delta^*$ is a necessary requirement in order to have particles with $p=p_{\max,0}(t)$ in the whole SNR.

It is interesting to note that under the assumption of test-particle DSA, where $\alpha= 3\sigma/(\sigma-1)$, the distribution function of confined particles as reported in Eq.~\eqref{eq:f_in(t,r)} becomes almost independent on $r$. In fact it results that $\epsilon=0$ and the function $\Lambda(t')$ has a very mild dependence on $r$. In such a case neglecting diffusion is justified because $\partial_r f_{\rm conf} \simeq 0$.

\subsection{Distribution of escaping particles}    
\label{sec:f_esc}
As soon as $t > t_{\rm esc}(p)$, particles with momentum $p$ cannot be confined anymore by the turbulence operating in the shock region and they start escaping. Note that in several works, the escape is treated as an instantaneous process, in the sense that all non-confined particles are assumed to be located outside the remnant right after $t_{\rm esc}(p)$, without accounting for the fact that the particles can still be propagating inside the SNR for some time. While this assumption can be considered a good approximation for studying the total particle spectrum released into the Galaxy, it is no more valid when attempting a description of the early phase of the escape process in the region close to the SNR, in particular in the estimate of the gamma-ray flux from that region. In fact, if the propagation outside of the SNR is diffusive, escaping particles have a finite probability to be scattered back and re-enter the SNR, even if they do not feel the shock discontinuity anymore and do not undergo any further acceleration. This process will be especially important if the level of turbulence in the vicinity outside of the SNR is much higher than the average Galactic one, in such a way that the confinement time in that region would be significantly enhanced. As discussed in \S~\ref{sec:intro}, there are several reasons to think that such an increase of the turbulence might be realized, including the CR self-generated turbulence. In Appendix~\ref{sec:turbulence} we show that such effect can, under certain conditions, be important especially in the close proximity of SNRs.

In order to describe the particle evolution at early times after the escape, namely for $t > t_{\rm esc}(p)$, an approximate solution is obtained by assuming that particles decouple from the SNR and their evolution is governed by pure diffusion. The particle evolution is hence described by the same Eq.~(\ref{eq:transport}) but dropping the terms including $u_{\rm sh}$, which gives
\begin{equation} 
\label{eq:transport_out}
 \frac{\partial{f_{\rm esc}}}{\partial{t}} = 
 \frac{1}{r^2} \frac{\partial}{\partial r} \left[ r^2 D(p) \frac{\partial{f_{\rm esc}}}{\partial{r}}\right] \,,
\end{equation}
where from now on we will address $f_{\rm esc}(t,r,p)$ as the distribution of non-confined particles. Since the particles will start escaping after they have been confined by the turbulence, this equation will be solved with an initial condition given by the distribution function of confined particles at $t = t_{\rm esc}(p)$. By defining $f_{\rm conf}(t_{\rm esc}(p), r, p) \equiv f_{\rm conf,0}(r,p)$, the initial condition reads as 
\begin{equation} 
\label{eq:initialC}
  \begin{cases} 
f_{\rm esc}(t_{\rm esc}(p), r,p) = f_{\rm conf,0}(r,p) \qquad  & r<R_{\rm sh}(t_{\rm esc}(p)) \\
f_{\rm esc}(t_{\rm esc}(p), r,p) = 0 \qquad  & \textrm{elsewhere} \,.
  \end{cases}
\end{equation}

\begin{figure*}
\centering
\includegraphics[width=0.48\textwidth]{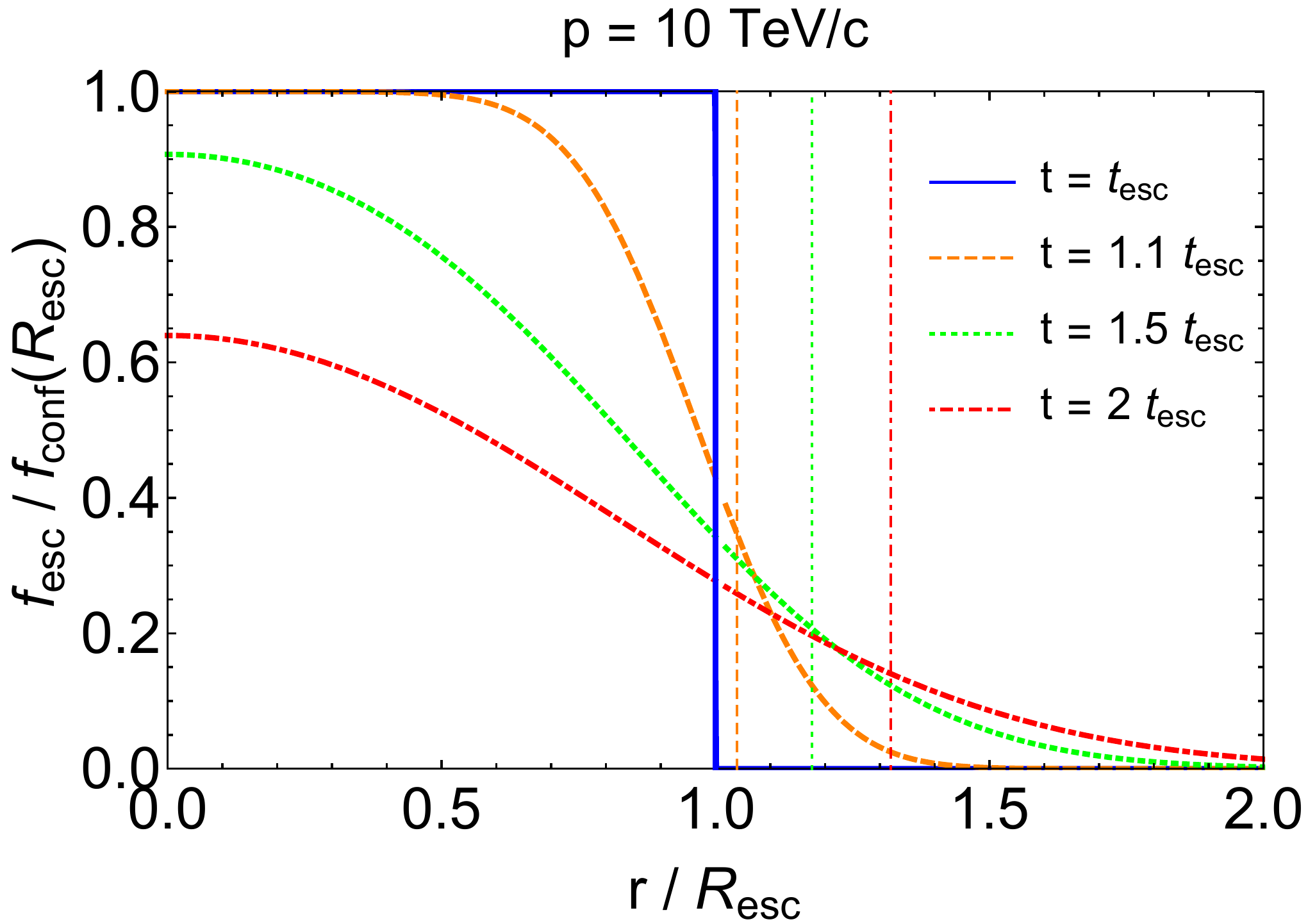}
\includegraphics[width=0.48\textwidth]{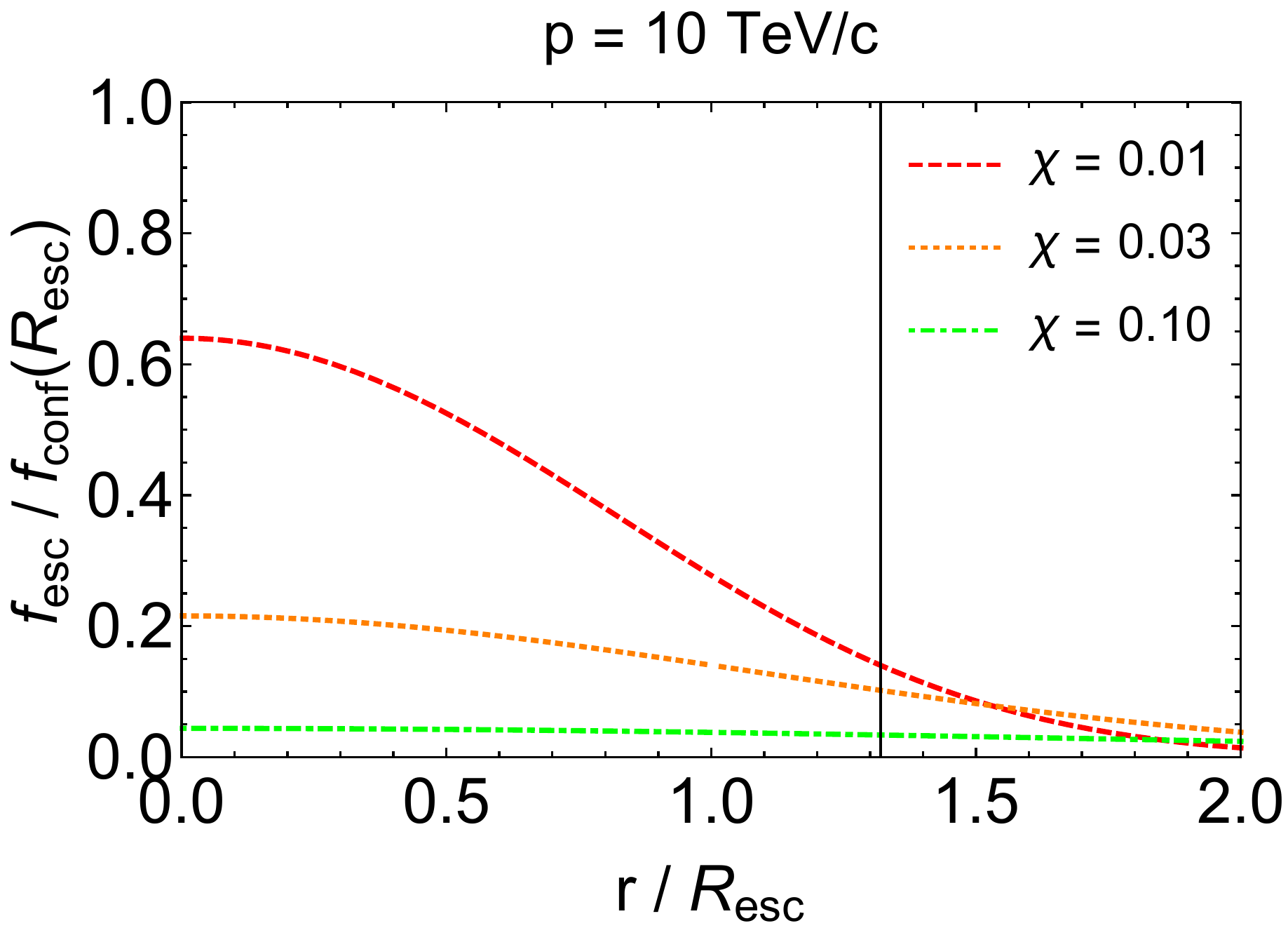}
\caption{Distribution of escaping particles as in Eq.~\eqref{eq:sol2}, at the arbitrary fixed momentum $p = 10$ TeV/c, as a function of the radial coordinate normalized to $R_{\rm esc}(p)$. For both panels we used $p_\textrm{M}=1$~PeV/c, $\delta = 4$ and $\alpha = 4$. {\it Left:} Different thick lines refer to different times, as labelled, and the vertical thin lines with the same color correspond to the shock position at those times. The diffusion coefficient is Kolmogorov-like, normalized to $\chi=0.01$. {\it Right:} Different lines refer to different value of the diffusion coefficient, as labelled. The time is fixed to $t=2 t_{\rm esc}$ and the vertical black line marks the shock position at that time.}
\label{fig:escape_alpha4}
\end{figure*}

\begin{figure*}
\centering
\subfigure[]{\label{fig:fescAlfa43} \includegraphics[scale=0.39]{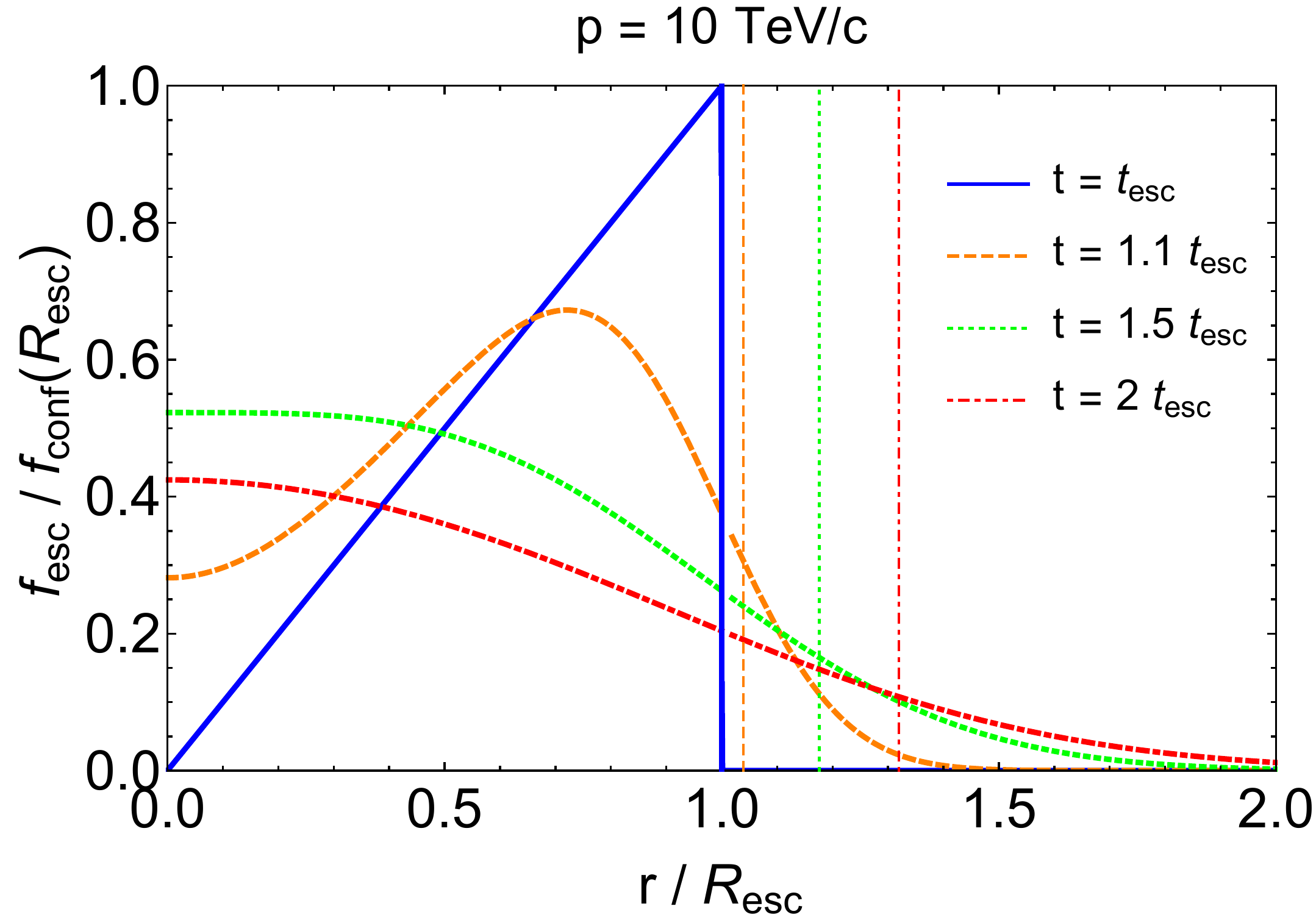}}
\subfigure[]{\label{fig:fescPalfa4} \includegraphics[scale=0.48]{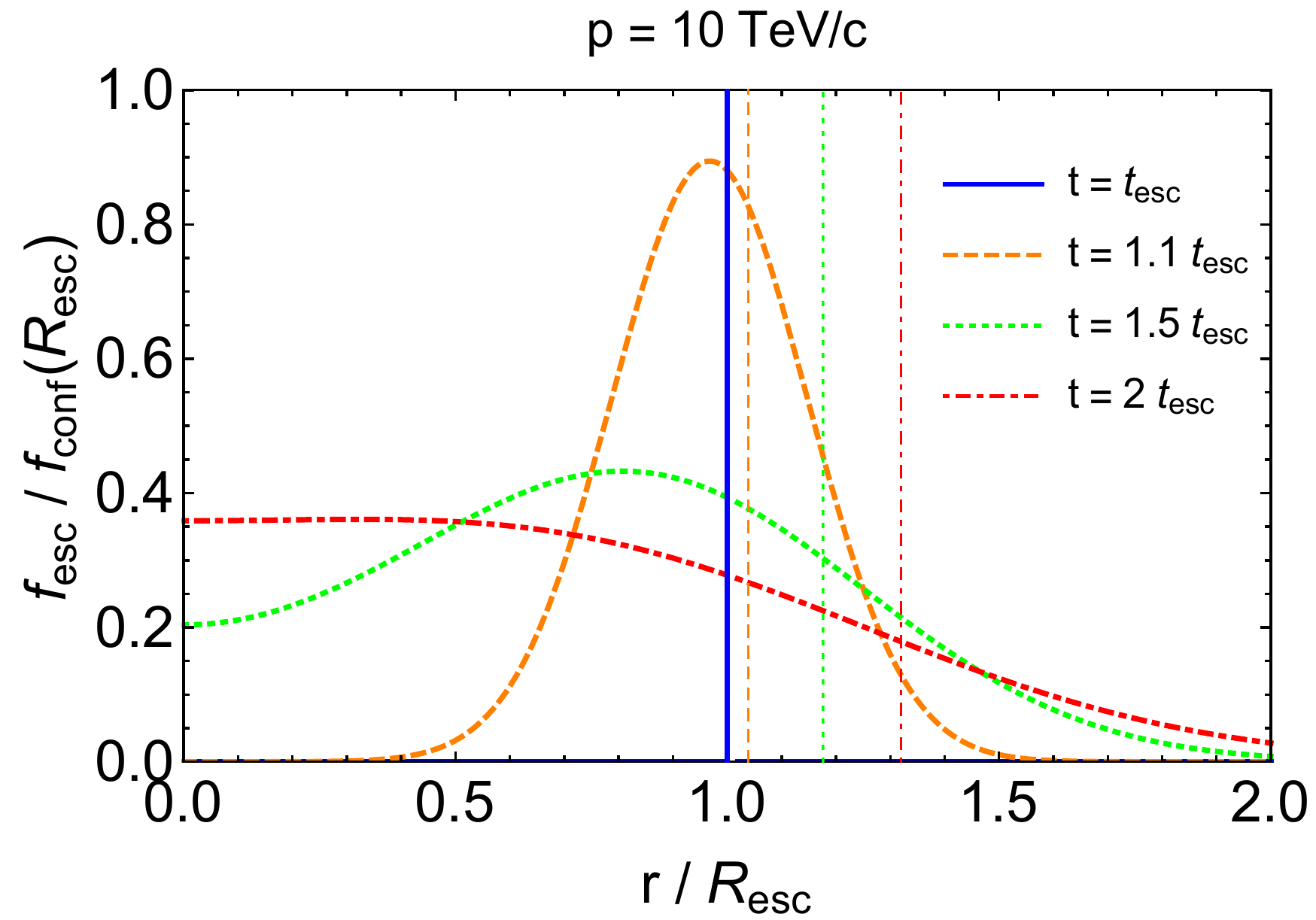}}
\caption{Distribution of escaping particles at the arbitrary fixed momentum $p = 10$ TeV/c, as a function of the radial coordinate normalized to $R_{\rm esc}(p)$. Different thick lines refer to different times, as labelled, and vertical thin lines represent the shock position at those times. The model here assumes $p_\textrm{max}(t)$ regulated by $p_\textrm{M}=1$~PeV/c and $\delta = 4$. The diffusion coefficient is Kolmogorov-like, normalized to $\chi = 0.01$.  {\it Left:} Escaping particles from the remnant interior, for an acceleration spectrum with slope $\alpha = 4+1/3$, as in Eq.~\eqref{eq:fesc43}.  {\it Right:} Escaping particles from the shock precursor, as in Eq.~\eqref{eq:precFinal}. }
\end{figure*}

The diffusion coefficient in the region outside the SNR, $D_{\rm out}$, is assumed to be spatially constant. Such assumption is made in order to derive an approximate analytic solution of Eq.~(\ref{eq:transport_out}), by using the method of Laplace transforms. As $D_{\rm out}$ is an unknown of the model, it might possibly be constrained by HE and VHE gamma-ray observations. Unless specified differently, we will assume a Kolmogorov-like diffusion, namely
\beq \label{eq:Dout}
  D_{\rm out}(p) \equiv \chi D_{\rm Gal}(p) = \chi 10^{28} \left( \frac{p c}{10 \,{\rm GeV}} \right)^{1/3} {\rm cm^{2} \, s^{-1}} \, ,
\eeq
where the parameter $\chi$ quantifies the difference with respect to the average Galactic diffusion coefficient.
Inside the SNR the diffusion coefficient $D_{\rm in}$ is in general different from the one outside, nevertheless for the sake of simplicity we will assume a homogeneous diffusion coefficient $D(p)$, such that $D_{\rm in}(p)= D_{\rm out}(p) \equiv D(p)$. Note that the analytical solution is only obtained for some values of the slope $\alpha$: we show here only two cases of interest, namely $\alpha=4$ and $\alpha = 4+1/3$, both assuming $\sigma=4$. The case $\alpha= 4$ corresponds to the standard case for DSA in the test-particle limit and the solution reads as (see Appendix~\ref{sec:appB} for the full derivation)
\begin{eqnarray} 
\label{eq:sol2}
  f_{\rm esc}(t,r,p) 
    = \frac{f_\textrm{conf,0}(p)}{2} \left\{ \frac{R_d}{\sqrt{\pi} \, r} \left[ e^{-\left( \frac{R_{+}}{R_d}\right)^2} - e^{-\left(\frac{R_{-}}{R_d}\right)^2} \right] + \right. \nonumber \\
    		\left. + {\rm Erf} \left( \frac{R_{+}}{R_d} \right) + {\rm Erf} \left( \frac{R_{-}}{R_d} \right) \right\} \times \Theta\left[ t - t_{\rm esc}(p) \right] \, ,
\end{eqnarray}
where $R_{\pm}(p) \equiv R_{\rm esc}(p) \pm r$,  $R_d (t,p) \equiv \sqrt{4 D(p) \left(t -t_{\rm esc}(p) \right)}$ is the diffusion length and  ${\rm Erf}(x)=2/\sqrt{\pi} \int_0^x e^{-z^2} dz$ is the error function. 
The spatial behavior of $f_{\rm esc}(r)$, as derived from Eq.~(\ref{eq:sol2}), is shown in Fig.~\ref{fig:escape_alpha4} for different times after the escape time and different normalizations $\chi$ of the diffusion coefficient.
Note that typical values of the parameters describing the evolution of a middle-aged SNR and the acceleration process have been adopted to obtain the plots, as indicated in Tab.~\ref{tab:tab1}. The results clearly show that, if $\chi$ is as small as $0.01$, roughly half of the escaped particles are still located inside the SNR at a time twice the escape time. \\
The second case considered, i.e. $\alpha=4+1/3$, represents a steeper acceleration spectrum, close to the values inferred from the gamma-ray observations of several SNRs (like Tycho and Cas~A) which have $\alpha \simeq 4.2\div4.3$. It is worth remembering that, to date, there is no consensus yet on the physical reason that would produce spectra steeper than $p^{-4}$. Some possibilities invoke the role of the speed of the scattering centers \citep{zirakashvili2008,morlinoCaprioli}, or the modification produced onto the shock structure by the presence of neutral hydrogen \citep{MB2016}, while a recent work ascribes the steepening to a combination of effects, including the shock spherical expansion, its temporal deceleration and the tilting of the magnetic field at the shock surface \citep{malkovAharonian2019}. Regardless of the physical reason producing such a steeper spectrum, we have chosen $\alpha = 4+1/3$ because an analytical solution for the non-confined particle density can be obtained, which is (see Appendix~\ref{sec:appB} for the full derivation)
\begin{equation}
\label{eq:fesc43}
\begin{split}
\frac{f_\textrm{esc}(t,r,p)}{k(t_\textrm{esc})} & = \left\{ \frac{R_d}{\sqrt{\pi}} e^{-\left(\frac{r}{R_d}\right)^2} + \frac{R_d}{2\sqrt{\pi}} \left( \frac{R_-}{r} \right) e^{-\left( \frac{R_+}{R_d} \right)^2} + \right. \\
& -  \frac{R_d}{2\sqrt{\pi}} \left( \frac{R_+}{r} \right) e^{-\left( \frac{R_-}{R_d} \right)^2}  +  \left( r+\frac{R_d^2}{2r} \right) \textrm{Erf}\left[ \frac{r}{R_d} \right] +  \\
& \left.  + \frac{1}{2} \left( r+\frac{R_d^2}{2r} \right) \textrm{Erfc}\left[ \frac{R_+}{R_d} \right] - \left( 1-\textrm{Erf} \left[ \frac{R_-}{R_d} \right] \right) \left( \frac{r}{2} + \frac{R^2_d}{4r} \right) \right\} \times \\
& \times \Theta[t-t_\textrm{esc}(p)] \, ,
\end{split}
\end{equation}
where ${\rm Erfc}(x)= 1-{\rm Erf}(x)$ and the function $k(t)$ reads as 
\begin{equation}
  k(t) = \frac{3 \xi_\textrm{CR} \rho_0}{25 \pi c (m_\textrm{p} c)^{4-\alpha} \Lambda(t)} 
  	\left( \frac{\xi_0 E_{\rm SN}}{\rho_0} \right)^{1/5} t^{-8/5}  	\,.
\end{equation}
The spatial behavior of the non-confined distribution function of Eq.~(\ref{eq:fesc43}) is plotted in Fig.~\ref{fig:fescAlfa43} for different times after $t_\textrm{esc}(p)$, where we fixed $p=10$ TeV/c and $\chi = 0.01$.  The main difference with respect to the solution presented in Eq.~(\ref{eq:sol2}) resides in the initial distribution function $f_{\rm conf}(r)$, which is flat in $r$ for the case $\alpha=4$, while it increases linearly with $r$ for $\alpha=4+1/3$. The difference at later times just reflects the different initial condition.


\subsection{The precursor region}    
\label{sec:precursor}
An additional contribution to the CR escaping density function comes directly from the shock precursor region \cite[see][]{Ptuskin+2005, Bell&Schure2014}. We can estimate such contribution by adopting the steady state solution of the transport equation in the plane shock approximation, which reads as
\begin{equation}
\label{eq:precExp}
f_\textrm{p} (t,r,p) = f_0(t,p) \exp \left[- \frac{u_{\rm sh}(t)}{D_\textrm{p}(p)} (r-R_{\rm sh}) \right] \, ,
\end{equation}
where $D_\textrm{p}(p)$ represents the diffusion coefficient within the precursor. In order to simplify the description of particle escaping from the precursor region, we approximate the exponential function in Eq.~\eqref{eq:precExp} with a $\delta$-function centered on the shock position $r=R_{\rm sh}$ such that it conserves the total number of particles contained in the precursor itself, namely
\begin{equation}
  f_\textrm{p,conf} (t,r,p) \simeq f_0(t,p)  \frac{D_\textrm{p}(p)}{u_{\rm sh}(t)} \delta (r-R_{\rm sh})  \,.
\label{eq:precDelta}
\end{equation}
As before, the temporal evolution of the particle density escaping the precursor region at $t>t_{\rm esc}(p)$ is described by Eq.~(\ref{eq:transport_out}), provided that Eq.~(\ref{eq:precDelta}) is adopted as its initial condition. The solution so obtained is found to be\footnote{Note that Eq.~\eqref{eq:precFinal} is formally identical to the solution found by \citet{ohira+2010} (their Eq.~(6)). Nevertheless, there is a fundamental difference between their work and ours: \citet{ohira+2010} assume that all particles accelerated by the SNR are located only at the shock when they start escaping, while we account for two different contributions, the one from the particle distribution inside the SNR (Eq.~\eqref{eq:sol2}) plus the one from the precursor. Both contributions are needed to correctly model the gamma-ray spectrum as explained in \S\ref{sec:gamma}.} (see Appendix~\ref{sec:appB})
\begin{equation}
\label{eq:precFinal}
\begin{split}
  \frac{f_\textrm{p,esc}(t,r,p)}{f_0(t_\textrm{esc},p)} = 
  & \frac{1}{\sqrt{\pi}} \frac{R_\textrm{esc}}{R_d} \frac{D_\textrm{p}(p)}{u_{\rm sh}(t_\textrm{esc})r} 
  	\left[ e^{-\left( \frac{R_-}{R_d}\right)^2} - e^{-\left( \frac{R_+}{R_d}\right)^2} \right] \times  \\	
  & \times \Theta[t-t_\textrm{esc}(p)]  \,.
\end{split}
\end{equation}
This is shown in Fig.~\ref{fig:fescPalfa4} as a function of the radial coordinate for $p= 10$ TeV/c and assuming $\chi = 0.01$ for the diffusion coefficient. The initial $\delta$-function rapidly expands, filling both the interior and the exterior of the remnant. For the chosen value of the parameters, at $t= 2 t_\textrm{esc}(p)$ the majority of the particles are still located inside the remnant also because the shock keeps moving. We will show in the next two Sections that the presence of particles escaped from the precursor and still located inside the shock radius can in principle produce a peculiar feature in the gamma-ray spectrum resulting from hadronic collisions occurring inside the remnant.

\section{The proton spectrum}    
\label{sec:protons}
The key result of this work is that the escape process can produce a particle spectrum inside the remnant different from the one accelerated at the shock. A general believe is that the escape should produce an exponential suppression of the particle spectrum at the highest energies. Nevertheless, if the diffusion coefficient is small enough, the contribution from non-confined particles in the remnant interior makes the final spectrum resembling rather a broken power-law distribution, as we will show in this Section, where we are going to derive the spectrum of particles located both inside and outside the SNR. \\

\begin{figure*}
\centering
\subfigure[]{ \label{fig:psAlfa4_D1e27} \includegraphics[scale=0.34]{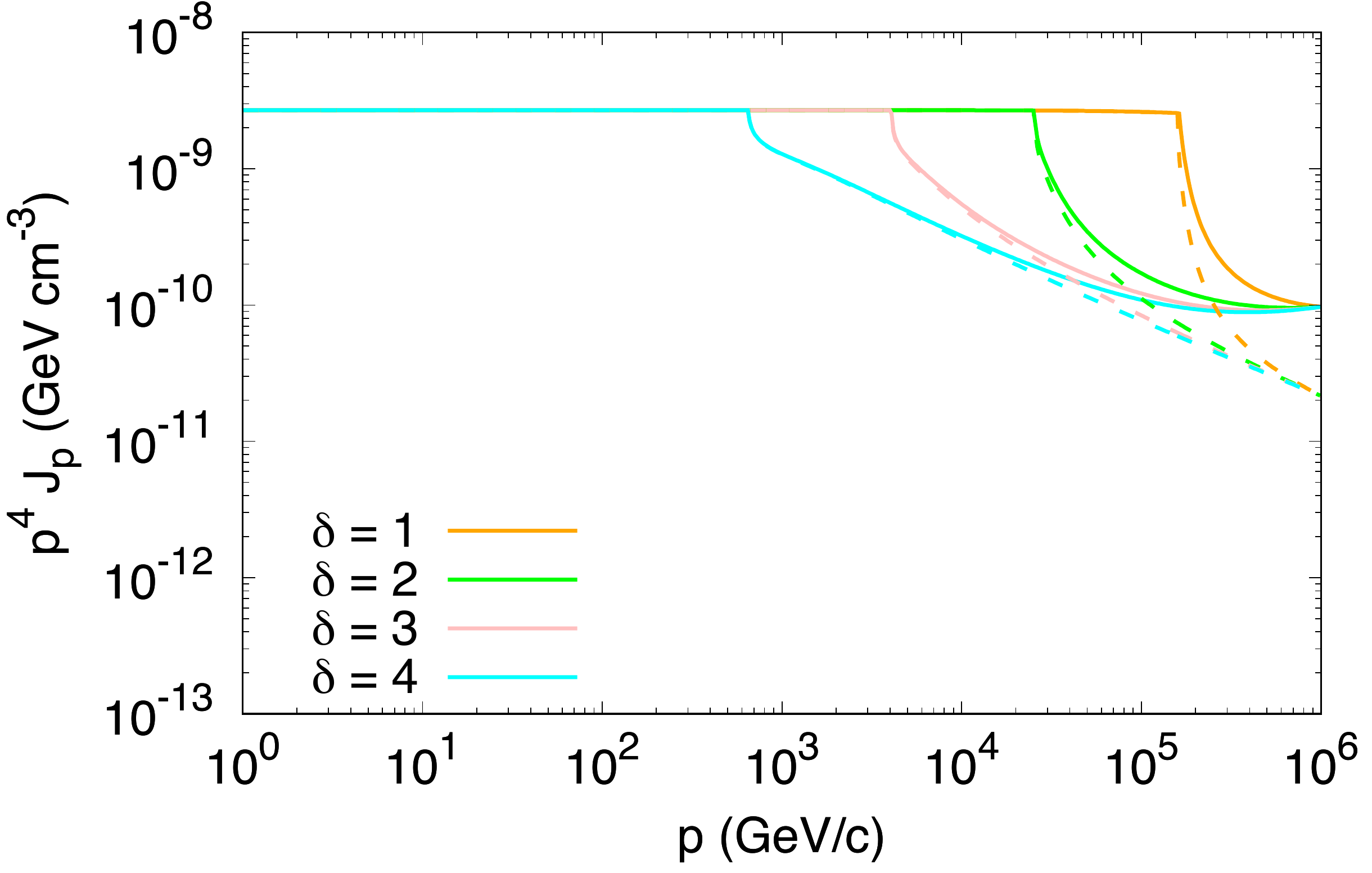}}
\subfigure[]{ \label{fig:psAlfa4_D1e28} \includegraphics[scale=0.34]{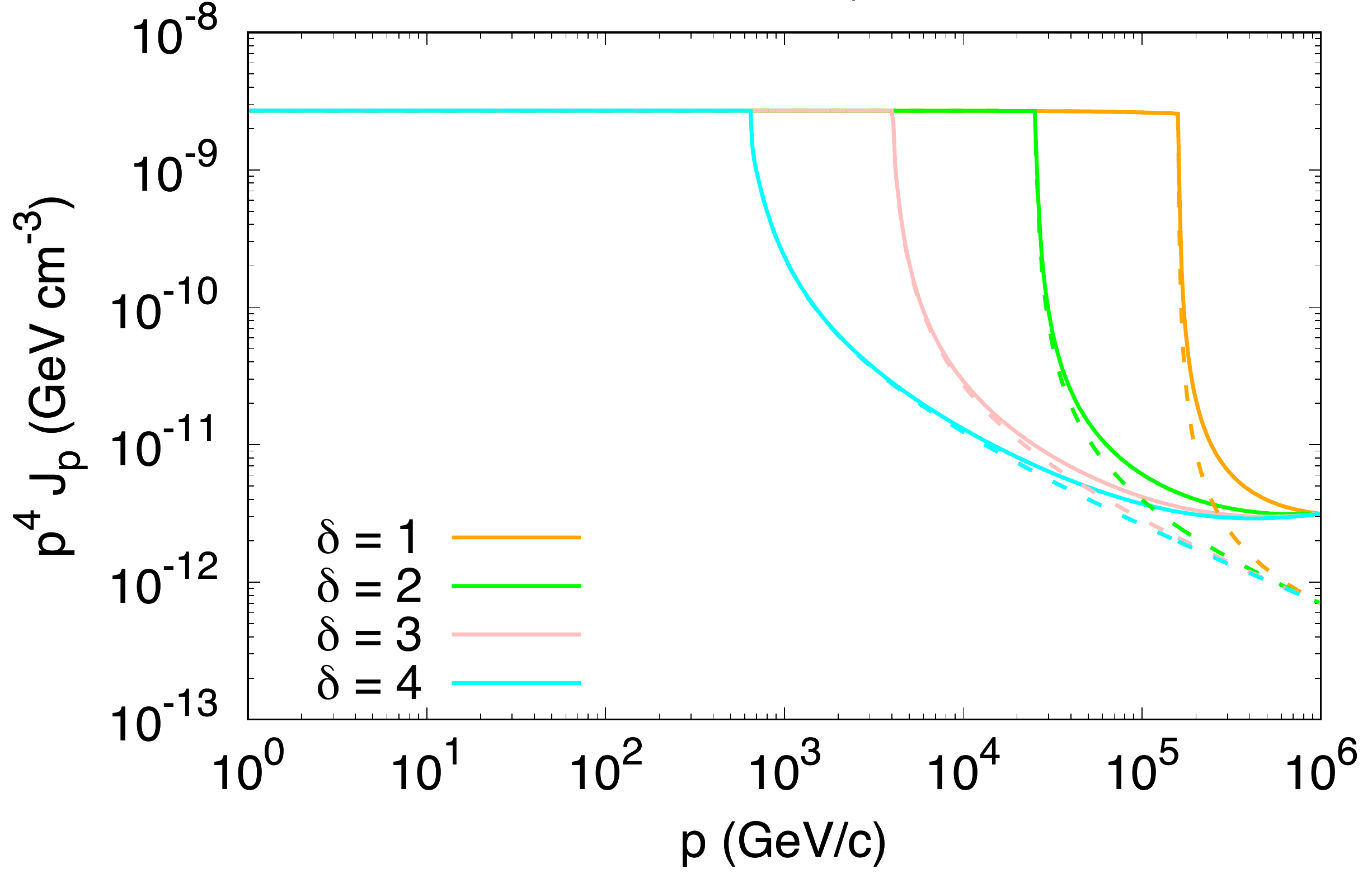}}
\caption{Total proton spectrum inside a SNR, contributed by confined plus non-confined particles. Different curves refer to different values of the index $\delta$, which regulates the time-dependence of the maximum momentum at the shock (see Eq.~\eqref{eq:pmax0}). The acceleration spectrum is assumed $\propto p^{-4}$ and the diffusion coefficient is normalized to $\chi=0.1$ (\emph{left panel}) and $\chi=1$ (\emph{right panel}). The remaining parameters are given in Tab.~\ref{tab:tab1} for both panels. The particle distributions without the contribution from shock precursor are always shown as dashed lines.}
\label{fig:psAlfa4} 
\end{figure*}

The average proton spectrum resulting from all the particles contained inside the remnant radius, including both confined and non-confined ones, as well as the contribution from particles released through time by the precursor, is computed as
\begin{eqnarray}
\label{eq:PSin}
  J_p^{\rm in} (t,p) = \frac{4\pi}{V_\textrm{SNR}} \int_0^{R_\textrm{sh}(t)}  
  		\left[ f_\textrm{esc}(t,r,p) + \hspace{2cm} \right.\nonumber \\
  	        \left. + f_\textrm{p,esc}(t,r,p)+f_\textrm{conf}(t,r,p) \right] r^2 dr \,,
\end{eqnarray}
where $V_\textrm{SNR}=4 \pi R^3_\textrm{sh}(t)/3$ is the remnant volume. The result of this computation is shown in Fig.~(\ref{fig:psAlfa4}), where we have assumed an acceleration spectrum $f_0 (p) \propto p^{-4}$ and a maximum momentum scaling with time given by Eq.~(\ref{eq:pmax0}) with $p_\textrm{M}=1$~PeV/c. Different values of the slope $\delta$ and of the diffusion coefficient normalization $\chi$ are explored, while the remaining parameters are fixed to the values given in Table~\ref{tab:tab1}.
\begin{table}
\centering
\footnotesize
\caption{Benchmark values for the set of parameters describing the SNR evolution and the particle acceleration: $E_{\rm SN}$ is the kinetic energy released at the SN explosion, $M_{\rm ej}$ the mass of the ejecta, $n_0$ the upstream density, $T_{\rm SNR}$ the remnant age, $\xi_{\rm CR}$ the acceleration efficiency and $p_\textrm{M}$ the maximum momentum at the Sedov time.}
\label{tab:tab1}
\begin{tabular}{cccccc}
\hline	
$E_{\rm SN}$  &  $M_{\rm ej}$  &  $n_0$  &  $T_{\rm SNR}$  &  $\xi_{\rm CR}$  & $p_\textrm{M}$ \\
\hline
$10^{51}$ erg  & 10 $M_{\odot}$ &  1 cm$^{-3}$  &  $10^4$ yr  & $10\%$  & 1 PeV/c \\
\hline  
\end{tabular}
\end{table}
%
%
%

\begin{table}
\caption{Values for the momentum break in the spectrum of protons confined within a middle-aged SNR, in the parametrization of Eq.~\eqref{eq:pmax0}. Benchmark values adopted from Tab.~\ref{tab:tab1}.}
\begin{center}
\begin{tabular}{cc}
\hline
$\delta$ & $p_\textrm{br}$~(GeV/c) \\ 
\hline 
$1$ & $1.6 \times 10^5$ \\
$2$ & $2.5 \times 10^4$ \\
$3$ & $4.1 \times 10^3$ \\
$4$ & $6.5 \times 10^2$ \\
\hline
\end{tabular}
\end{center}
\label{tab:pmaxT}
\end{table}

In all the plotted spectra a break is clearly visible at $p_{\rm br} = p_{\max,0}(T_{\rm SNR})$, that is the maximum momentum achieved in correspondence of the shock position at the observation time (the remnant age). Its value depends on the parameter $\delta$ which regulates how fast the maximum momentum decreases with time, as shown in Tab.~\ref{tab:pmaxT}. Below and above the break the spectrum is contributed by confined and non-confined particles, respectively. At any given time, the spectral trend above the momentum break strongly depends on $\delta$ and on the energy dependence of the diffusion coefficient assumed. On the other hand, the number of particles contributing above the break is regulated by the normalization value of the diffusion coefficient: by comparing Figs.~\ref{fig:psAlfa4_D1e27} and \ref{fig:psAlfa4_D1e28}, where respectively $\chi=0.1$ and $\chi=1$ were adopted, one can derive that by increasing the value of the diffusion coefficient, the amount of non-confined particles located inside the SNR is reduced and the spectral break rather becomes a sharp cut-off. In addition, a flattening is visible at the highest energies, where the contribution of particles escaping prom the precursor becomes important. In fact the spectrum of particles contained in the precursor is harder than that of particles located inside the SNR, being proportional to $f_0(p) D_\textrm{p}(p)$.

\begin{figure*}
\centering
\subfigure[]{\label{fig:psCardillo} \includegraphics[scale=0.34]{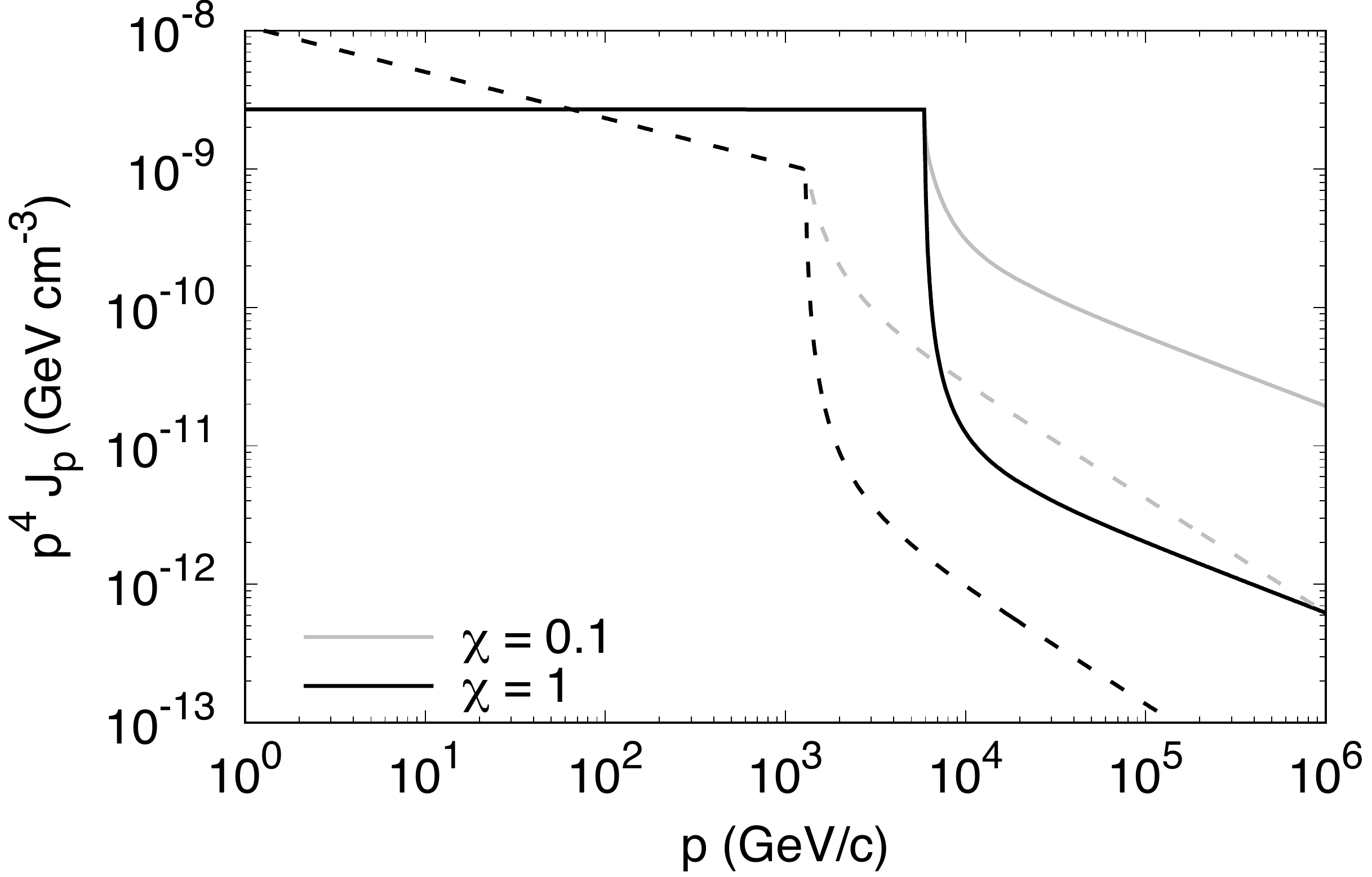}}
\subfigure[]{\label{fig:protonsOut} \includegraphics[scale=0.34]{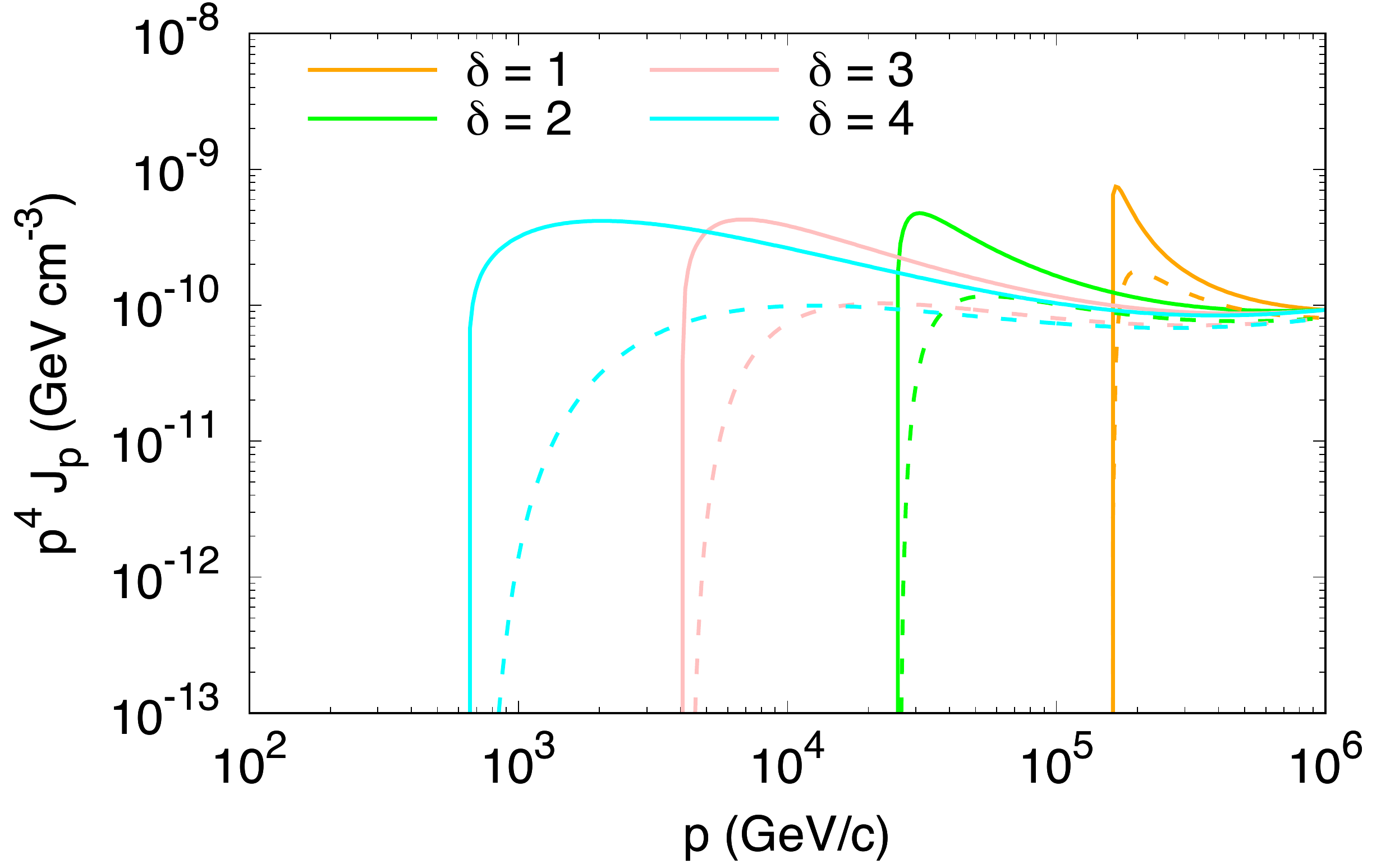}}
\caption{\textit{Left:} Total proton spectrum inside a SNR, calculated by adopting the time evolution of the maximum momentum as in \citet{Cardillo+2015}: solid lines refer to acceleration slope $\alpha=4$, while dashed ones refer to $\alpha=4+1/3$. The diffusion coefficient is normalized with $\chi=0.1$ for grey lines and $\chi=1$ for black lines.  \textit{Right:} Spectrum of non-confined protons located outside of the remnant shell at $T_\textrm{SNR}=10^4$~yr and for different values of the slope $\delta$ as labelled. The particle spectrum is integrated inside a spherical corona extending either between $R_{\rm sh}(T_{\rm SNR})$ and $2 R_{\rm sh}(T_{\rm SNR})$ (solid lines) or between $2R_{\rm sh}(T_{\rm SNR})$ and $3 R_{\rm sh}(T_{\rm SNR})$ (dashed lines). The diffusion coefficient is normalized to $\chi = 0.1$. The remaining parameters are given in Tab.~\ref{tab:tab1} for both panels.}
\end{figure*}

It is now worth comparing the results obtained above with the case of a different recipe for the time dependence of the maximum energy at the shock. We will use the calculation from \citet{Cardillo+2015} as summarized in \S~\ref{sec:pmax} (Eqs.~\eqref{eq:emaxCardillo4} and \eqref{eq:tescCardillo4} for the case with $\alpha=4$, and Eqs.~\eqref{eq:emaxCardillo43} and \eqref{eq:tescCardillo43}  for the case with $\alpha=4+1/3$). In this scenario, by adopting the same parameter values as in Fig.~\ref{fig:psAlfa4_D1e27}, we obtain a systematically softer spectrum above the break with respect to what was obtained with the power-law dependence of $p_\textrm{max}$. Concerning the energy break, in the scenario described by \citet{Cardillo+2015} we derive $p_\textrm{br} (\alpha=4) \simeq 5.9 \times 10^3$~GeV/c and $p_\textrm{br} (\alpha=4+1/3) \simeq 1.3 \times 10^3$~GeV/c. The results are reported in Fig.~\ref{fig:psCardillo}, for the two aforementioned values of the acceleration spectrum slope $\alpha$.

The spectrum of protons located outside of the SNR includes only non-confined particles. Considering a spherical corona between the radii $R_1$ and $R_2$ (with $R_{\rm sh} \leq R_1 < R_2$), the average spectrum is given by
\beq
\label{eq:PSout}
  J_p^{\rm out} (t,p) = \frac{3}{R_2^3-R_1^3} \int_{R_1}^{R_2}  
  		\left[ f_\textrm{esc}(t,r,p) + f_\textrm{p,esc}(t,r,p) \right] r^2 dr \,.
\eeq
Such a spectrum is shown in Fig.~\ref{fig:protonsOut} for two positions of $R_1$ and $R_2$, where a Kolmogorov-like diffusion coefficient normalized to $\chi = 0.1$ is assumed. A low-energy threshold is visible at $p=p_\textrm{br}$, while the peak of the distribution is regulated by the amount of particles with propagation length equal to the radial extension of the corona, i.e. $R_\textrm{d}(p) \approx R_2-R_1$. The contribution from the precursor is well visible at the highest energies, where the spectrum flattens like in the case shown in Fig.~\ref{fig:psAlfa4}. The different line styles refer to different spatial integration regions: solid lines refers to a corona between $R_{\rm sh}(T_{\rm SNR})$ and $2 R_{\rm sh}(T_{\rm SNR})$, while dashed lines are spectra calculated for particles located between $2 R_{\rm sh}(T_{\rm SNR})$ and $3 R_{\rm sh}(T_{\rm SNR})$. It can be noted that, towards the outer regions of the accelerator, the low-energy cut-off of the spectrum is moved to highest energies since only the most energetic particles can reach the farther regions. As a consequence, also the spectrum normalization is affected, and it decreases moving outwards. The peculiar bump-like shape of the primary spectrum in the external regions of the shock implies that, in the presence of a dense target of gas, the secondary radiation resulting from hadronic collisions will show a similar feature. It is thus timely to investigate the expected gamma-ray emission connected with hadronic collisions of accelerated protons both within the shock radius and outside of it, in order to understand whether next-generation instruments could be able to detect the VHE gamma-ray halos possibly surrounding SNRs as generated by escaping particles. In fact, an SNR population study might shed light on how diffusion operates in these sources and even provide information on how the escape process works, by constraining the slope of the maximum momentum with time from a statistical point of view. \\

\section{Gamma rays from hadronic collisions}    
\label{sec:gamma}
In this Section we will evaluate the gamma-ray flux resulting from hadronic collisions occurring both inside and outside a middle-aged SNR, by calculating {\it i}) the volume integrated emission in the remnant itself, {\it ii}) the volume integrated emission in different annular regions immediately outside the shock radius, and {\it iii}) the projected radial profile, which is an extremely relevant information when dealing with extended objects. Note that the shock of a middle-aged remnant is expected to expand outside of the wind termination shock, but still inside the cavity of hot and rarified medium blown by the stellar progenitor \citep{castor,dwa2005}. In such a region, the medium has a homogeneous density, as we will consider in the following. 

\subsection{Volume integrated emission} 
\label{sec:gamma_volume}
We have shown in \S~\ref{sec:protons} that a characteristic energy break appears in the spectrum of protons contained in the SNR interior right at the maximum momentum that particles achieve through the shock acceleration process at the SNR age. Analogously, the spectrum of secondaries resulting from proton collisions with the target gas (the so-called pp interaction) will reflect this feature. For a remnant expanding into a homogeneous medium with number density $n_0=\rho_0/m_{\rm p}$, the density profile of the plasma $n_{\rm in}$, that is expanding with the SNR evolving during the ST phase, can be well approximated by the following polynomial expression
\begin{equation}
\label{eq:sedovDensity}
  n_{\rm in}(t,r) = n_0 \sigma \left[ a_1 X^{\alpha_1} + a_2 X^{\alpha_2}  + a_3 X^{\alpha_3} \right] \, ,
\end{equation}
where $X= r/R_{\rm sh}(t)$. The parameters in Eq.~\eqref{eq:sedovDensity} have been derived by fitting the radial density profile of the SNR interior as presented in \citet{sedovBook}, thus obtaining the following values: $a_1=0.353$, $a_2=0.204$, $a_3=0.443$, $\alpha_1=4.536$, $\alpha_2=24.18$ and $\alpha_3=12.29$. 

\begin{figure*}
\centering
\subfigure[]{\label{fig:gammaInAlfa4} \includegraphics[scale=0.33]{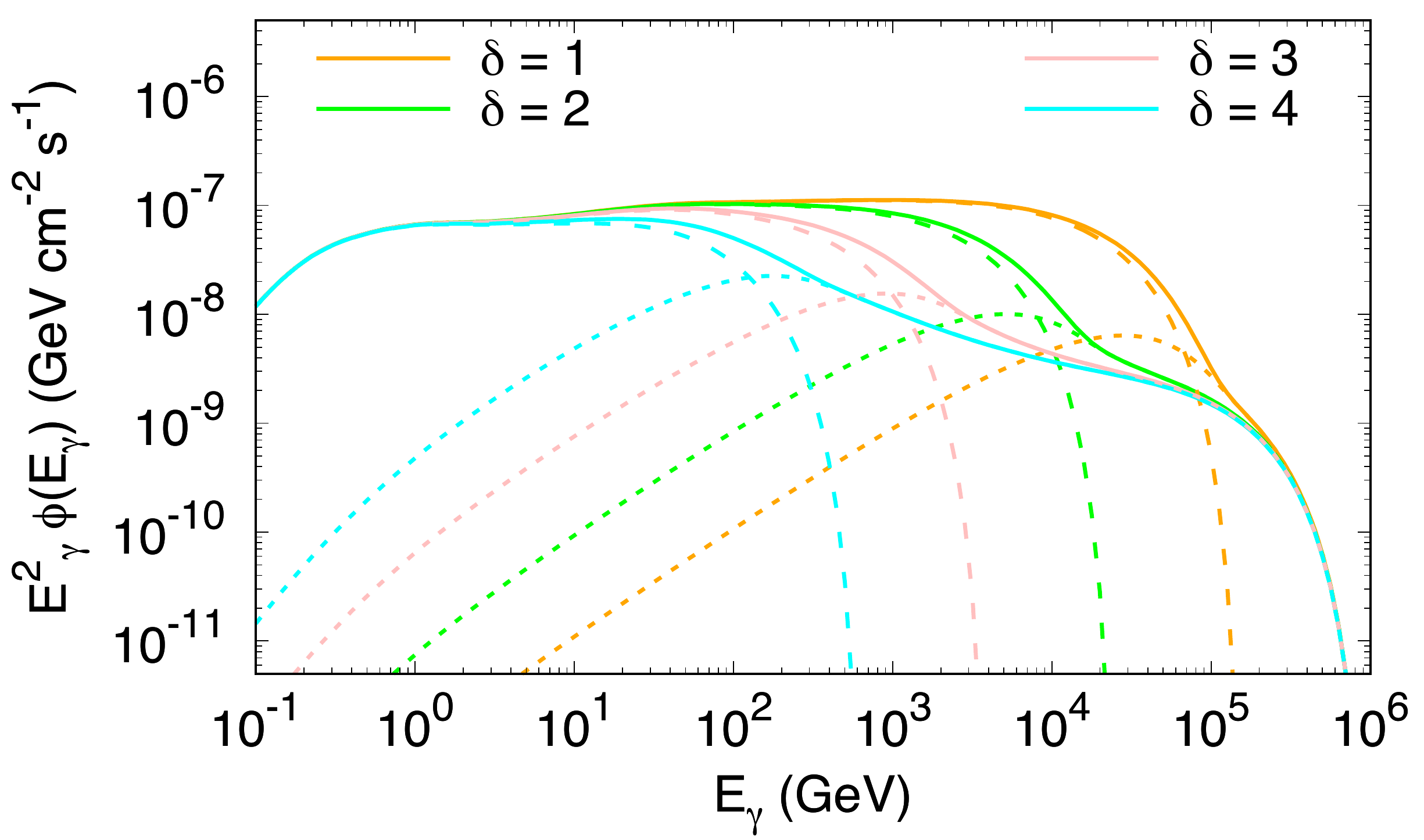}}
\subfigure[]{\label{fig:gammaOut}\includegraphics[scale=0.33]{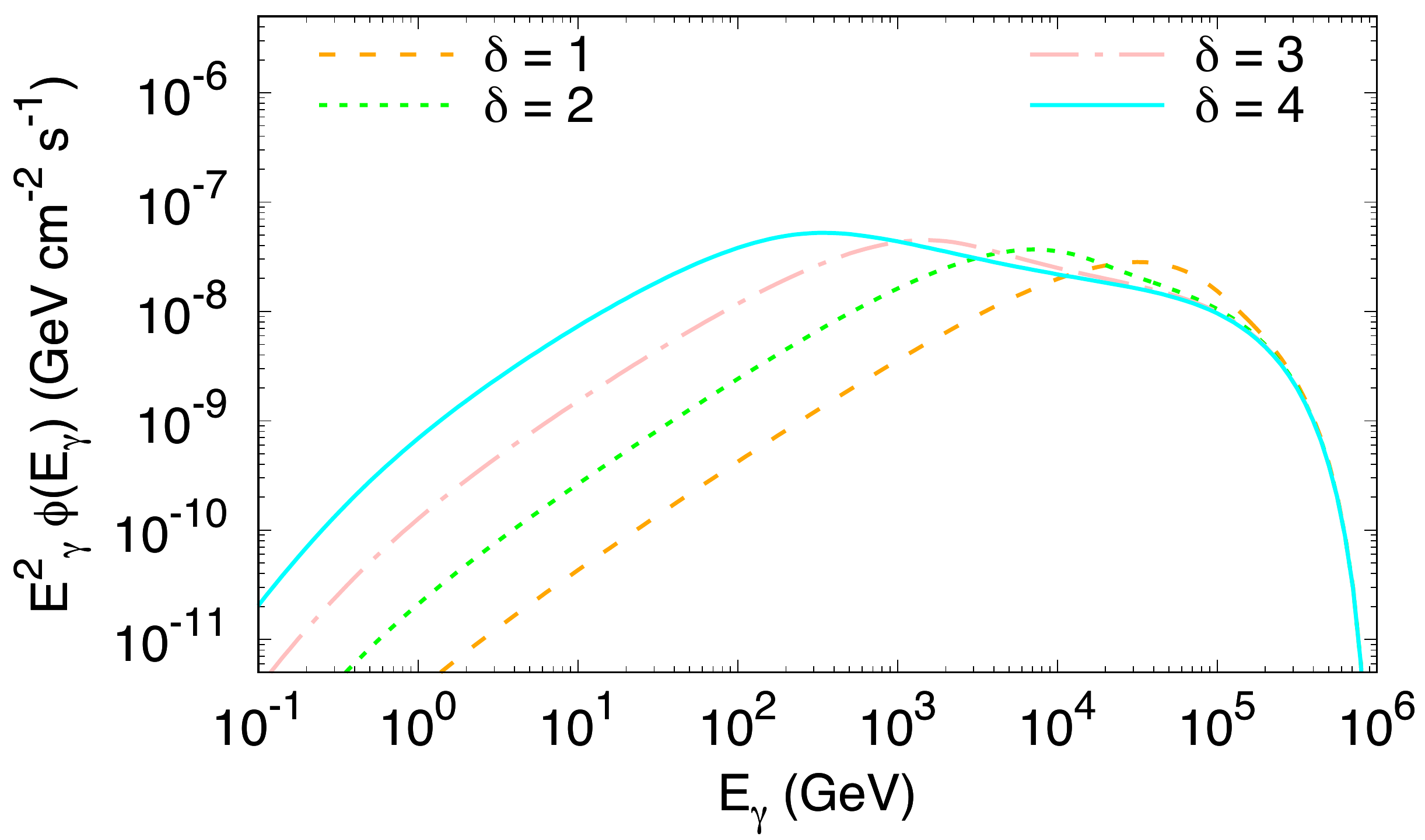}}
\caption{Gamma-ray flux from hadronic collisions in a middle-aged remnant SNR located at a distance of $d=1$~kpc. The acceleration spectrum has been fixed with slope $\alpha=4$ and the diffusion coefficient normalized to $\chi=0.1$. The maximum momentum temporal dependence has been parametrized according to Eq.~\eqref{eq:pmax0}, for different values of $\delta$, as labelled. The remaining parameters are given in Tab.~\ref{tab:tab1} for both panels. \textit{Left:} Emission by confined (dashed lines) and non-confined particles (dotted lines) located inside the SNR, where solid lines refer to the sum of the two contributions.  \textit{Right:} Emission from escaped particles located in an annulus extending from $R_\textrm{SNR}$ to $2R_\textrm{SNR}$ outside of the SNR.}
\label{fig:gamma}
\end{figure*}

\begin{figure}
\centering
\includegraphics[scale=0.33]{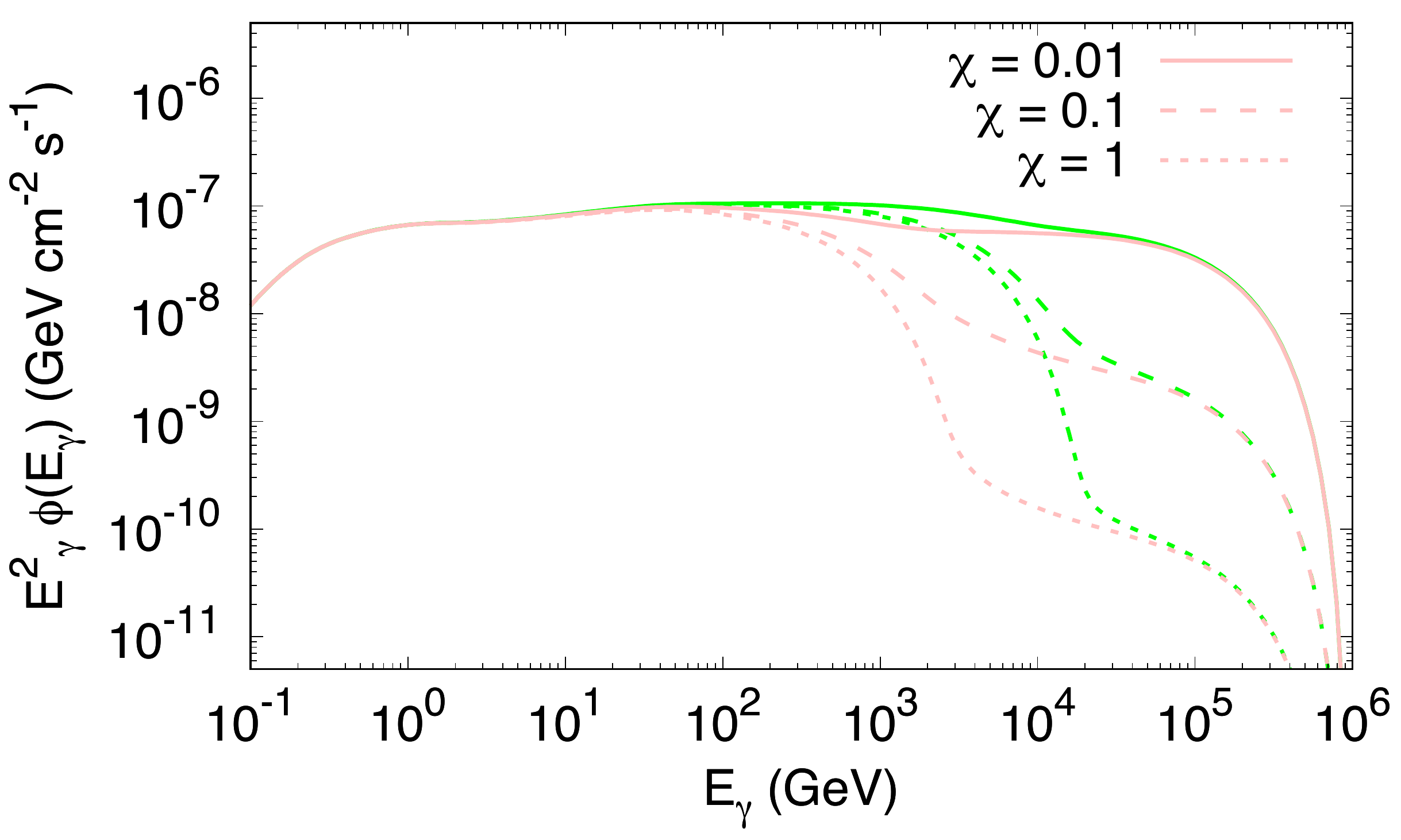}
\caption{Gamma-ray flux from hadronic collisions in a middle-aged SNR located at a distance of $d=1$~kpc. The acceleration spectrum has been fixed with slope $\alpha=4$, while the maximum momentum temporal dependence has been parametrized according to Eq.~\eqref{eq:pmax0}, with $\delta=2$ (green lines) and $\delta=3$ (pink lines). Different normalizations of the diffusion coefficient are explored, as labelled. The remaining parameters are given in Tab.~\ref{tab:tab1} for both panels.}
\label{fig:gammaInAlfa4D} 
\end{figure}

Convolving the differential energy spectrum of protons residing in the remnant interior with the density profile of Eq.~\eqref{eq:sedovDensity}, and considering the differential cross section for pp collisions, one derives  
the differential energy flux of secondary gamma rays expected at different times. We parametrized the differential cross section following \cite{Kafexhiu+2014}, adopting the parameter values they obtained from SYBILL 2.1. The resulting gamma-ray flux is shown in Fig.~\ref{fig:gamma}, where an SNR located at a distance of $d=1$~kpc is considered. The two panels show the integrated emission from the SNR interior (Fig.~\ref{fig:gammaInAlfa4}) and from a spherical corona around it (Fig.~\ref{fig:gammaOut}). The same parameter values as in Fig.~\ref{fig:psAlfa4_D1e27} have been used. The gamma-ray spectrum reflects the behavior of the proton distribution, namely the emission from the remnant interior shows a break at an energy $\sim 0.1 p_{\rm br} c$ that depends on the value of $\delta$, in that also the break in the proton spectrum depends on it, as shown in Tab.~\ref{tab:pmaxT}. Below the break the flux is dominated by confined particles, while above it is due only to the non-confined particles. The relative intensity of the two contributions, and hence the shape of the transition, is primarily determined by the diffusion coefficient $D_{\rm out}$ as can be seen in Fig.~\ref{fig:gammaInAlfa4D}: the smaller the diffusion coefficient, the larger the confinement time, implying that a larger amount of particles will be still residing within the remnant at a fixed remnant age. On the other hand, $D_{\rm out}$ does not affect the emission from the confined particles, as expected from Eq.~\eqref{eq:f_in(t,r)}. \\

It is worth to discuss here few aspects of the model. In the presence of massive gas clouds embedded in the shock environment, the CR propagation might result affected \citep{celli2018} and consequently a simple rescaling of the gamma-ray emission with the gas density does not apply. On the other hand, the main conclusion derived here, namely the presence of a break in the gamma-ray spectrum due to escaping CR, is not affected by a possible time dependence of CR acceleration efficiency (provided the dependence is smooth), though quantitative results may change. In particular, the slope of the gamma-ray spectrum beyond the break is expected to become harder (softer) if $\xi_{\rm CR}$ is a decreasing (increasing) function of time. \\

It is interesting to note that high-energy observations point towards the presence of a break in the spectrum of middle-aged SNRs, like W~44 and IC~443 \citep{FermiLAT2013}, located respectively at $22 \pm 8$~GeV and $279 \pm 34$~GeV. To this respect, the recent results by \cite{Zeng-Xin-Liu:2019} are of special interest: using a spectral fitting procedure, the authors inferred the presence of a break in the gamma-ray spectrum of the majority of SNRs in a sample of $\sim 30$ objects. The energy break is observed to decrease with the remnant age, ranging from $\sim 10$ TeV for younger SNRs (age of $\sim 10^3$ yr) down to few GeV at ages of few $10^4$ yr. This is compatible with our assumption of a maximum energy which decreases in time and assuming a slope $\delta$ roughly in between 2 and 3, which is needed to reproduce the spectral break observed in the gamma-ray spectrum at few tens of GeV for $T_{\rm SNR} \simeq 10^4$ yr. Note that a more quantitative constraint on the value of $\delta$ requires a detailed analysis of each individual SNRs in the sample, accounting for a correct evaluation of their evolutionary stage, the density and spatial distribution of the circumstellar medium, the possible presence of IC emission, as well as the presence of PWN associated with the remnants. In a forthcoming paper we will apply our model to few selected middle-aged SNRs, in order to derive constraints on the time dependence of particle escape  as well as on the diffusion coefficient in the circumstellar region. \\

The gamma-ray spectrum emitted from a coronal region outside the SNR between $R_{\rm sh}$ and $2 R_{\rm sh}$ is shown in Fig.~\ref{fig:gammaOut}, corresponding to the proton spectrum shown in Fig.~\ref{fig:protonsOut}. The photon emission peaks at $E_{\rm peak} \simeq 0.1 \hat{p} c $, where $\hat{p}$ is the momentum of particles that at $t= T_{\rm SNR}$ have reached the external boundary of the corona and, hence, have completely filled this region. $E_{\rm peak}$ ranges from $\sim 100$ GeV up to tens of TeV for the chosen values of the parameters.

It is worth stressing that a distinctive signature of the escape scenario, as presented in this work, is that the break energy of the spectrum from the SNR interior is tightly connected to the peak energy of the spectrum from the outside regions. Next-generation gamma-ray instruments, as CTA, would possibly investigate such connection in middle-aged SNRs. Nonetheless, a correct evaluation of the instrument performances requires to account for the spatial extent of the region under investigation: e.g., a remnant with age $T_\textrm{SNR}= 10^4$~yr at a distance $d=1$~kpc would cover an angular area of radius $\sim 0.8$~deg, resulting into an even more extended halo of escaping particles. Because the large amount of background coincident with such large angular search window tends to degrade the instrument sensitivity level \citep{ambrogi18}, it is likely that only bright Galactic emitters will show gamma-ray fluxes large enough to explore both the contribution from inside the shock radius and that from the closer outer regions.

\subsection{The gamma-ray radial profile}
\label{sec:gamma_profile}
\begin{figure*}
\centering
\subfigure[]{\includegraphics[scale=0.3]{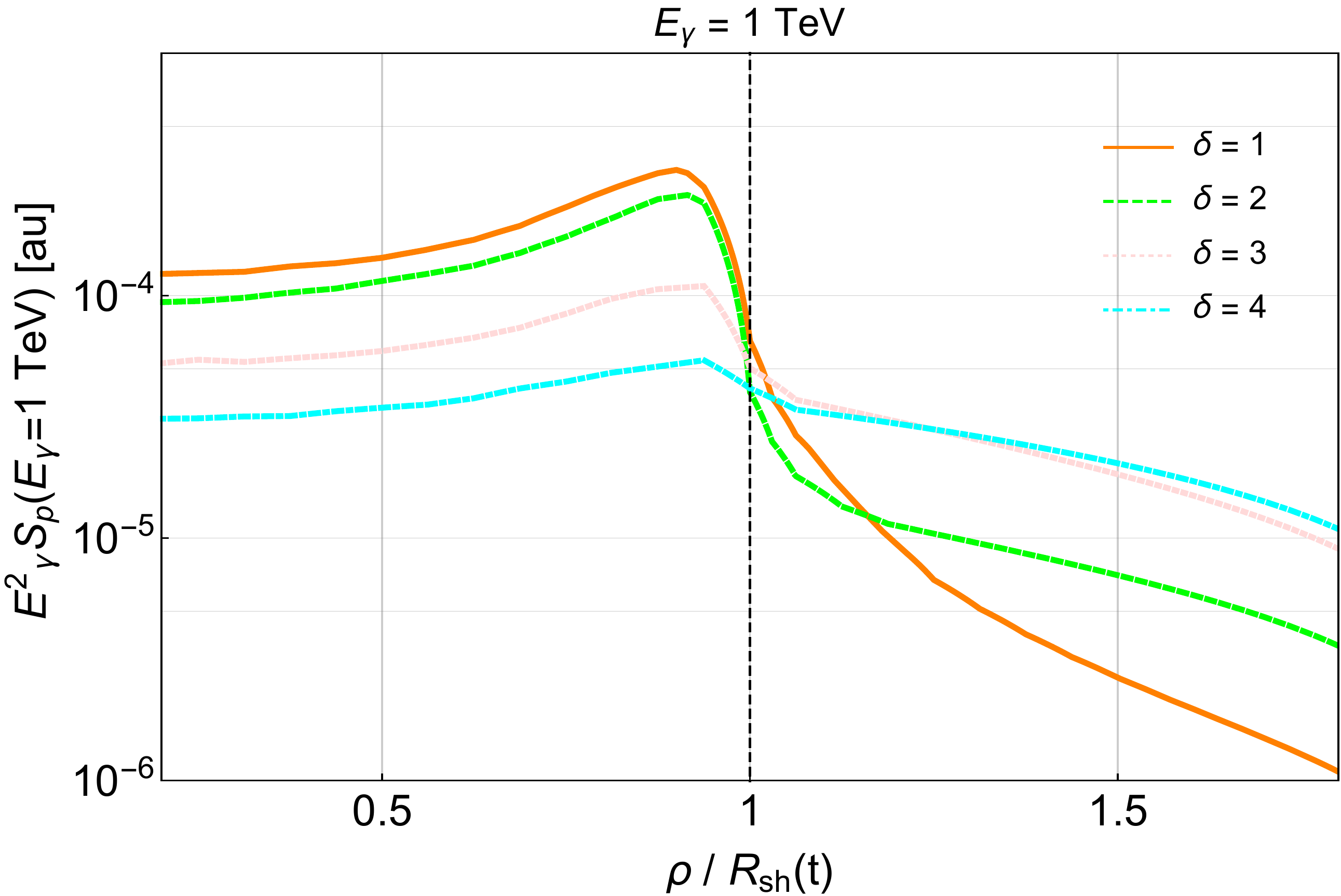}}
\hspace{0.5cm}
\subfigure[]{\includegraphics[scale=0.315]{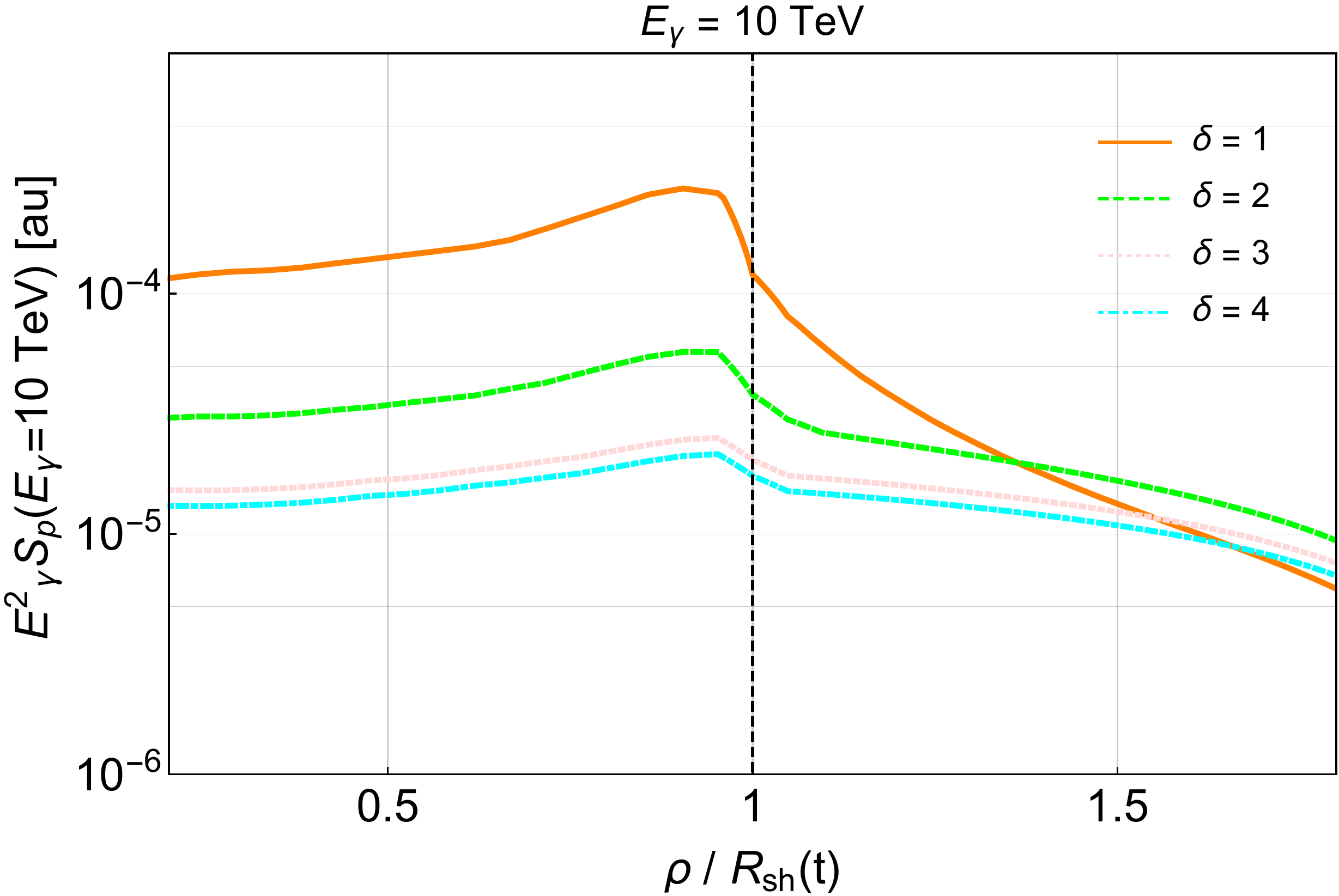}}
\caption{Radial profile of the (a) $1$~TeV and (b) $10$~TeV gamma-ray surface brightness projected along the line of sight according to Eq.~\eqref{eq:gamma_emissivity} with $R_{\max}=2R_\textrm{sh} (t)$. Hadronic collisions are considered in a middle-aged SNR located at a distance of $d=1$~kpc. The acceleration spectrum has been fixed with slope $\alpha=4$, while the diffusion coefficient is normalized to $\chi=0.1$. The maximum momentum temporal dependence has been parametrized according to Eq.~\eqref{eq:pmax0}, with slope $\delta$ as labelled. The remaining parameters are given in Tab.~\ref{tab:tab1} for both panels. The vertical black line represents the shock position. Arbitrary units are adopted for the surface brightness. }
\label{fig:gammaProf}
\end{figure*}

The volume-integrated emission is not always the best quantity to compare with the observations if the object under exam is spatially extended. In this case, precious information can be derived from the remnant morphology, especially from the radial profile of the emissivity. In order to compare the observed radial profiles with the model predictions, one has to project the radial emission along the line of sight $l$. Under the assumption of spherical symmetry, the spatial dependence of the gamma-ray emissivity at energy $E_\gamma$ can be summarized uniquely through its radial dependence, $S(E_\gamma, t, r)$. As a consequence, the projected emission expected at a distance $\rho$ from the remnant center, namely the surface brightness $S_{\rm p}(E_{\gamma},t,\rho)$, can simply computed by integrating the radial emission along the line of sight, as
\beq  
\label{eq:gamma_emissivity}
  S_{\rm p}(E_{\gamma},t,	\rho) = 2 \int_0^{\sqrt{R_{\max}^2 - \rho^2}}  
  	S(E_\gamma, t, r=\sqrt{\rho^2 + l^2}) dl
\eeq
where $R_{\max}$ defines the radial extension of the region considered in the projection. Fig.~\ref{fig:gammaProf} provides an example of the expected gamma-ray surface brightness profile arising from pp interactions at different photon energies, namely at $E_{\gamma} = 1\,{\rm TeV}$ and $E_{\gamma} = 10\,{\rm TeV}$, and for different slopes $\delta$. Here, the instrumental performances are also accounted for, in that a Gaussian smearing of the angular resolution is applied to the profile model of Eq.~(\ref{eq:gamma_emissivity}). A point spread function with values of $\sigma (E_{\gamma} = 1\,{\rm TeV}) = 0.051^{\circ}$ and $\sigma (E_\gamma = 10\, {\rm TeV}) = 0.037^{\circ}$ is adopted in the following, as these represent the performances that next generation of imaging atmospheric Cherenkov telescopes, as CTA\footnote{\href{https://www.cta-observatory.org/science/cta-performance/}{https://www.cta-observatory.org/science/cta-performance/}}, are expected to achieve. A drop of the surface brightness is visible beyond the shock position, as expected in the case of shell-like SNRs. However, the jump strongly depends on the value of $\delta$, in that it appears that the larger is $\delta$ the smaller is the jump: this is connected with the fact that a faster decrease in the maximum momentum temporal dependence implies that even low-energy particles have escaped the shock and populate the region beyond the remnant shell. Moreover, the drop appears to shrink with increasing photon energy, as parent particles are able to reach larger distances. As the emission profile drop ranges from about one to two orders of magnitude, it appears likely that the next-generation instruments will achieve the sensitivity level necessary for detecting such an emission from outside of the shell of bright emitters.

\section{The CR spectrum injected into the Galaxy}    
\label{sec:finj}
In the last years, the escape problem has received much attention by several authors \cite[]{Caprioli+2010a, ohira+2010, Drury2011, Malkov+2013, Bell&Schure2014, Cardillo+2015}. However, because this process depends on several subtleties of the acceleration process, there is not yet a consensus about what is the most realistic approach to model it. \citet{ohira+2010} found that the spectrum of run-away particles during the Sedov stage can be both softer and harder than that at the acceleration site, depending on the assumptions for the injection process as well as the spectrum of accelerated particles. In particular, under the condition that the CR acceleration efficiency is constant in time, they found that a particle spectrum that is accelerated at the source flatter than $E^{-2}$ will result in an $E^{-2}$ escape spectrum, whereas a steeper acceleration spectrum will result in an escape spectrum with equal steepening. This result was also obtained by \citet{Bell&Schure2014}, who clearly formulated it by using a more physically motivated framework which links the escaping process to the level of magnetic field amplification, under the same assumption that a fixed fraction of energy is transferred to CRs. Later \citet{Cardillo+2015} confirmed the result, and discussed its implications in the context of maximum energy achievable in both type Ia and type II SNe. A different assumption was considered by \citet{brose2019}, who used a time-dependent code to compute the accelerated spectrum assuming a constant fraction of particle injected into the accelerator. In such a case they obtained a spectrum of escaping particles steeper than $p^{-4}$ (in the hypothesis that the diffusion coefficient at the shock is self generated). In this Section, we will calculate the total particle spectrum released by a single SNR evolving during the ST phase, according to the modeling developed in \S~\ref{sec:model}. \\
Since we assumed that for $t > t_{\rm esc}$ particles are completely decoupled from the SNR evolution, the total density of CR with momentum $p$ injected into the Galaxy by an individual SNR is given by the integral of all particles contained inside the radius of the SNR at the time of escape, i.e.
\begin{equation} 
\label{eq:f_inj}
  f_{\rm inj}(p) = 4\pi  \int_{0}^{R_{\rm esc}(p)} r^2 f_{\rm conf} \left(t_{\rm esc}(p), r, p \right) dr \,.
\end{equation}
In this expression we omitted the contribution due to particles located in the precursor ahead of the shock. In fact, this contribution can be neglected if one assumes that the diffusion coefficient inside the precursor is much smaller than $R_{\rm esc}(p) u_{\rm sh}(t_{\rm esc})$ at all times $t < t_{\rm esc}(p)$. 

The confined density function $f_{\rm conf}$ was given in Eq.~\eqref{eq:f_in(t,r)}, where the spatial dependence was hidden in $t'(t,r)$, given by Eq.~\eqref{eq:t_sh}. Using also Eq.~\eqref{eq:ST_R}, the following relation is derived
\begin{equation}
  \frac{t'(t_{\rm esc},r)}{t_\textrm{esc}} = \left[\frac{r}{R_{\rm esc}(p)} \right]^{5 \sigma/2}
\end{equation}
so that the confined function at the escape time can be expressed as
\begin{equation}
  f_\textrm{conf}(t_\textrm{esc},r,p) = f_0 ( t_{\rm esc},p) \, R_\textrm{esc}^{\alpha(\sigma-1)-3} 
  	\frac{\Lambda(p)}{\Lambda(p_{\max,0}(t'))}  \,.
\end{equation}
Introducing this expression into Eq.~\eqref{eq:f_inj}, one obtains
\begin{equation}
\label{eq:finj2}
  f_{\rm inj}(p) = 4\pi f_0(t_\textrm{esc}(p),p) R^{3}_\textrm{esc}(p)
	                  \int_0^{1} y^{\alpha(\sigma-1)-1} \frac{\Lambda(t)}{\Lambda(t')} dy \, ,
\end{equation}
where we recall that $\Lambda(t)$ is a shortcut for $\Lambda(p_{\max, 0}(t))$.
As explained in \S~\ref{sec:f_shock}, the acceleration spectrum produced at every time is assumed to scale as a fixed fraction of the ram pressure (see Eq.~\eqref{eq:f_0}), namely $f_0(p) \propto u_\textrm{esc}^2(p) p^{-\alpha}/\Lambda(p)$, where $u_\textrm{esc}(p) =u_\textrm{sh}(t_\textrm{esc}(p))$. Therefore
\begin{equation}
  f_\textrm{inj}(p) \propto \frac{u_\textrm{esc}^2(p) R^3_\textrm{esc}(p)}{\Lambda(p)} p^{-\alpha} \mathcal{I}(p) \, ,
\end{equation}
$\mathcal{I}(p)$ being the integral in Eq.~\eqref{eq:finj2}.
Under the assumption that $t_\textrm{esc}(p) \propto p^{-1/\delta}$ (see Eq.~\eqref{eq:tesc}) and that the remnant is undergoing the ST phase, then $R_\textrm{esc}(p) \propto p^{-2/5\delta}$ and $u_\textrm{esc}(p) \propto p^{3/5\delta}$. Hence, the momentum dependence enclosed in the term $u_\textrm{esc}^2(p)$ perfectly balances that of $R^3_\textrm{esc}(p)$, and the spectrum injected in the Galaxy is simply given by
\begin{equation}
\label{eq:finj1}
f_\textrm{inj}(p) \propto \frac{\mathcal{I}(p)}{\Lambda(p)} \, p^{-\alpha} \,.
\end{equation}
Neglecting the dependency on particle momentum provided by $\mathcal{I}(p)/\Lambda(p)$, one would derive that the the spectrum injected in the Galaxy coincides with the acceleration spectrum. But, the inclusion of these additional terms makes the solution more involved. The integral $\mathcal{I}(p)$ reduces to a pure number in both the relativistic and the non-relativistic limit, but it has a tiny dependence on $p$ for trans-relativistic energies. $\Lambda(p)$, on the contrary, reads in the relativistic limit ($p \gg m_{\rm p} c$) as $\propto p_{\min}^{4-\alpha}$ for $\alpha > 4$ and $\propto p^{4-\alpha}$ for $\alpha < 4$. In the non-relativistic limit, however, $\Lambda (p) \propto p_{\min}^{5-\alpha}$ $\alpha >5$ and $\propto p^{5-\alpha}$ for $\alpha < 5$, respectively. Given these limits, we derive that the injected spectrum is, for $p \gg m_{\rm p} c$,
\begin{equation}
\label{eq:finj1final}
f_\textrm{inj}(p) \propto  
\begin{dcases}
p^{-\alpha} \, \qquad \alpha>4 \\
p^{-4} \, \qquad \alpha<4 \, ,\\ 
\end{dcases}
\end{equation}
while for particles with $p \ll m_pc$ it rather holds
\begin{equation}
\label{eq:finj2final}
f_\textrm{inj}(p) \propto 
\begin{dcases}
p^{-\alpha} \, \qquad \alpha>5 \\
p^{-5} \, \qquad \alpha<5 \, .\\ 
\end{dcases}
\end{equation}

\begin{figure}
\centering
\includegraphics[width=0.48\textwidth]{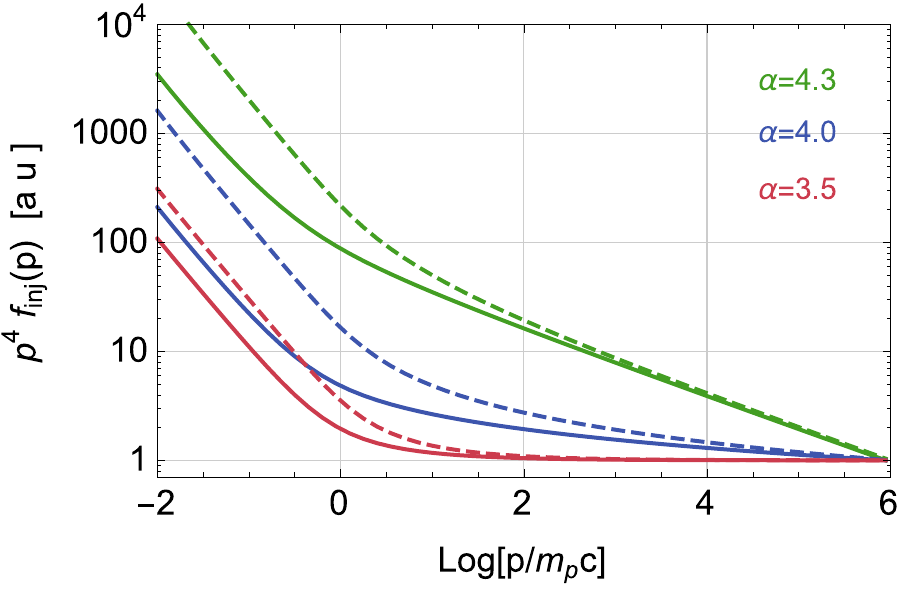}
\caption{Spectrum injected into the Galaxy for three different cases of acceleration spectrum $f_0(p) \propto p^{-\alpha}$ with $\alpha=4.3, 4.0$ and 3.5 (solid lines from top to bottom). The corresponding dashed lines show the approximate solution given by $p^{-\alpha}/\Lambda(p)$ for the same values of $\alpha$. The maximum momentum temporal dependence has been parametrized according to Eq.~\eqref{eq:pmax0}, with slope $\delta=3$.}
\label{fig:injected-spectra} 
\end{figure}

\noindent
In summary, for particles with $p \gg m_{\rm p} c$, we find a result analogous to what already found by \citet{ohira+2010}, \citet{Bell&Schure2014} and \citet{Cardillo+2015}, namely: {\it i}) if the acceleration spectrum is steeper than $p^{-4}$, the spectrum injected in the Galaxy will show the same steepness, thus coinciding with the acceleration spectrum; {\it ii}) if the acceleration spectrum is flatter than $p^{-4}$, the spectrum injected in the Galaxy will be a $p^{-4}$ power law, regardless of the acceleration spectrum. This behavior is shown in Fig.~\ref{fig:injected-spectra} where we compare $p^{-\alpha}/\Lambda(p)$ versus $p^{-\alpha} \mathcal{I}(p)/\Lambda(p)$. The inclusion of the function $\mathcal{I}(p)$ does not modify the asymptotic behavior of $f_{\rm inj}$, as it only shifts the transition towards smaller energies. Note that, similarly to the implications derived for the gamma-ray spectrum of an individual SNR, the time-dependency of $\xi_{\rm CR}$ might modify the final spectrum released in the Galaxy: if $\xi_{\rm CR}$ decrease (increase) with time, then the injected spectrum results harder (softer).

At this point it is worth stressing that the result obtained in Eq.~\eqref{eq:finj1final} for relativistic energies coincides with past calculations by \citet{schure2013,Bell&Schure2014} and \citet{Cardillo+2015}, obtained under the same assumption that a fixed fraction of the shock energy is transferred to CRs. However, the definition of escaping particles adopted here is different from what has been assumed in the cited works. In fact, in \citet{schure2013,Bell&Schure2014} and \citet{Cardillo+2015}, the escaping spectrum at time $t$ is modeled as a $\delta$-function in energy which carries a fixed fraction of the kinetic energy that the shock has at the same moment $t$, namely $E_\textrm{esc} \propto \rho_0 u^2_\textrm{sh}(t)$. On the contrary, in the model presented here, the escaping flux at each fixed time $t$ includes particles that have been accelerated in the past when the shock speed was faster than $u_\textrm{sh}(t)$, and have also suffered adiabatic losses. In other words, the energy carried by the particles escaping at time $t$ is not a fixed fraction of $\rho_0 u^2_\textrm{sh}(t)$. The definition used by \citet{schure2013,Bell&Schure2014} and \citet{Cardillo+2015} is probably more suitable to describe the escape process during the initial phase of the remnant life. Nonetheless, the results obtained in the relativistic regime are consistent with each other. 

The result for non-relativistic energies, as expressed in Eq.~\eqref{eq:finj2final}, predicts a spectral steepening for $p \ll m_{\rm p} c$ if $\alpha < 5$. This result is at odd with the observed CR spectrum \citep{Cummings+2016}, where a hardening is rather observed. The disagreement is not surprising, in that two strong assumptions were set, which likely are not realized in reality: {\it i}) the shock keeps accelerating particles always maintaining the same efficiency, and {\it ii}) the remnant evolution proceeds all the way through the ST stage.

A consistent description of the particle spectrum injected in the Galaxy requires to account for the moment when the shock stops accelerating particles. Such a condition could be fulfilled when the SNR transits towards the snowplough phase or even before it, e.g. if the shock impacts on a neutral cloud where the ion-neutral friction destroys the magnetic turbulence, making it impossible for particles to keep diffusing around the shock.

In any case, the end of the acceleration could produce some kind of signature in the injected spectrum. We recall that the observed CR spectrum is a straight power-law in momentum down to $\sim$ few GeV, where a hardening is observed, but it is usually attributed to Galactic propagation effects, rather than processes occurring at the accelerator. Another interesting feature has been identified in the Voyager~1 data \citep{Cummings+2016}, where a hardening of the spectrum is observed at $E \lesssim 200$ MeV. Such a hardening cannot be explained only through ionization losses in the ISM, but it is rather the slope of the injection spectrum that has to be range from $p^{-4.3}$ to $\sim p^{-3.75}$ respectively for protons and Helium (with heavier elements showing a harder trend) \citep{Tatischeff-Gabici2018}.

\begin{figure}
\centering
\includegraphics[width=0.48\textwidth]{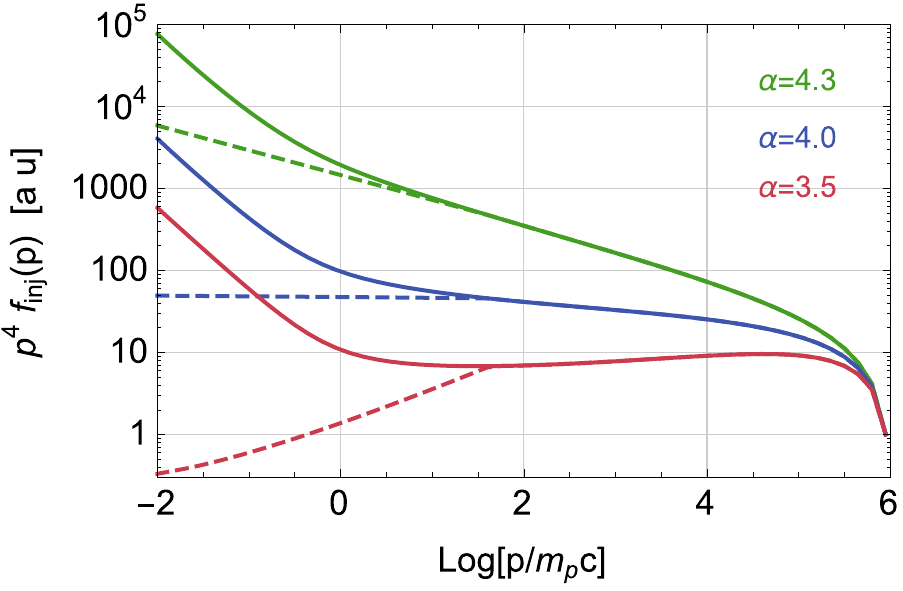}
\caption{Spectrum injected into the Galaxy for $\delta=3$ and $\alpha=4.3$, 4.0 and 3.5 (from top to bottom) with an increasing maximum energy during the free expansion phase. Dashed lines show the result when the acceleration is stopped at $t= 50 t_{\rm Sed}$ and particles still inside the SNR are instantaneously released.}
\label{fig:injected-spectra-delta} 
\end{figure}

Given the complex phenomenology of observations, it is worth investigating more in details the effect of the end of the acceleration mechanism on the particle spectrum injected in the Galaxy. To this purpose, we will calculate the spectrum produced by a SNR assuming that the acceleration suddenly stops at the beginning of the snowplow phase, which is reached at a time $t_{\rm sp}$ when the temperature of the shocked gas drops below $10^6$~K. For the parameter values assumed in Tab.~\ref{tab:tab1}, this condition is realized when $u_{\rm sh} \lesssim 200$ km s$^{-1}$, corresponding to an age of $t_{\rm sp} = 47$ kyr. The resulting $f_{\rm inj}$ is shown in Fig.~\ref{fig:injected-spectra-delta} and compared with the case of endless acceleration (solid lines). Three possible values of $\alpha$ are assumed, while $\delta$ is fixed to 3. When the acceleration stops, all those particles still located inside the SNR (namely with $p < p_{\max,0}(t_{\rm sp}) \simeq 40$ GeV/c) are instantaneously released into the ISM without suffering further adiabatic losses. As a consequence, the spectrum below 40 GeV/c is $\propto p^{-\alpha}$ and, interestingly, if $\alpha < 4$ a break appears right at this energy, while for $\alpha > 4$ the final spectrum does not show any feature. The case $\alpha = 4$ is somewhat border line, because the slope at high energies is slightly steeper than 4 (i.e. $\sim 4.1$) and a small spectral break is still visible.

In summary, the spectrum injected into the Galaxy is a featureless power law under two conditions: {\it i}) the acceleration spectrum has to be steeper than $p^{-4}$, and {\it ii}) the acceleration should stop when the maximum energy is still in the relativistic domain. The latter condition also translates into an upper bound for $\delta$ which, for the parameter values adopted here, has to be $\lesssim 4$.

A final comment concerns the cut-off present in the spectra shown in Fig.~\ref{fig:injected-spectra-delta}. Such a cut-off is not due to the shock acceleration process, which we assumed to produce a straight power law up to its maximum momentum, but is rather due to the increase of the maximum energy for times smaller than $t_{\rm Sed}$ (see Eq.~\eqref{eq:pmax0}). In Fig.~\ref{fig:injected-spectra}, on the contrary, the cut-off is absent because we assumed that the relation $p_{\max,0} \propto t^{-\delta}$ holds even for $t \leq t_{\rm Sed}$ to better show the asymptotical behavior of the spectra.

\section{Discussion and conclusions} 
\label{sec:conc}
The deviation observed in the HE and VHE gamma-ray spectra of SNRs, particularly in middle-aged ones, with respect to the simple spectral shape predicted by the DSA theory, might possibly be connected to the particle escape from the shock region. The escape process of particles accelerated at SNR shocks remains one of the less understood pieces of the shock acceleration theory. As a consequence, this aspect is often neglected, though it represents a fundamental part of the process, needed to explain the CR spectrum observed at Earth. In this paper, we presented a phenomenological model for the description of particle escape from a SNR shock aimed at evaluating the effects produced by the escape process on the spectrum of particles contained in the remnant and those located immediately outside of the shock region. In particular, when particles are not confined any more by the shock, they start to freely diffuse in the CSM, eventually escaping the accelerator. 
Within the assumption that the particle diffusion in the region outside of the remnants is suppressed with respect to the average Galactic diffusion, the escape process is not instantaneous and a relevant fraction of high-energy particles can still be located inside the SNR or close to it even once they are not confined anymore by the shock turbulence, producing diffuse gamma-ray halos around the remnant. 
Note that a one-dimensional anisotropic diffusion model could mimic the effect of a suppression of the diffusion coefficient  \citep[see, e.g.,][]{Nava-Gabici2013} and perhaps the spectral break discussed in this paper could result even without requiring a strong suppression of the external diffusion.

The escape process has at least two important consequences on the gamma-ray emission: {\it i}) the spectrum from the SNR interior observed at a fixed time presents a steepening above the maximum energy of particles accelerated at that time, and {\it ii}) the spectrum emitted from the halo around the remnant shows a low-energy cut-off at the energy corresponding to that of the escaping particles at the remnant age. While the second aspect could be tested with future gamma-ray telescopes (for instance CTA), the former could have already been detected. 

In fact, several SNRs show a spectral break in the gamma-ray spectrum. This founding has been summarized in a recent paper by \citet{Zeng-Xin-Liu:2019}, who showed that the majority of SNRs in a sample of $\sim 30$ objects presents evidence for a spectral break. Interestingly, the energy break decreases with increasing age, ranging from $\sim 10$ TeV for younger SNRs (age of $\sim 10^3$ yr) down to few GeV at ages of few $10^4$ yr. This result is in agreement with our interpretation of the energy break as due to the escape process where the maximum energy decreases like $E_{\max} \propto t^{-\delta}$ with $\delta$ between 2 and 3. Note, however, that the result found by \citet{Zeng-Xin-Liu:2019} should be taken as indicative, in that in order to derive a more reliable constraint a careful analysis object-by-object is needed to account for the correct evolutionary stage as well as a possible role of leptonic contribution. To this extent, in a forthcoming paper we will apply our model to some selected cases of middle-aged SNRs, in order to derive constraints on both the particle escape process and the diffusion coefficient in the circumstellar region of each specific remnant.

Finally, the total CR spectrum injected into the Galaxy by an individual SNR, evolving in the ST phase, has been computed. For an acceleration spectrum $\propto p^{-\alpha}$, under the assumption that a fixed fraction of the shock kinetic energy is converted into accelerated particles at every time, the spectrum injected into the Galaxy by a single SNR turns out to be: {\it i}) at $p \gg m_{\rm p} c$, $f_\textrm{inj}(p) \propto p^{-4}$ if $\alpha<4$ or $f_\textrm{inj}(p) \propto p^{-\alpha}$ if $\alpha>4$, and {\it ii}) at $p \ll m_{\rm p}c$, $f_\textrm{inj}(p) \propto p^{-5}$ if $\alpha<5$ and $f_\textrm{inj}(p) \propto p^{-\alpha}$ if $\alpha>5$. This result is independent on the temporal behavior of the maximum energy at the shock, but it relies on the assumption that the acceleration never stops and that the SNR always evolves in the ST phase. Furthermore, we also showed that the final injected spectrum can be a straight power law in momentum if the acceleration stops before the remnant enters the snowplough phase and if the slope is $\alpha > 4$. If these two conditions are not fulfilled, in general a spectral change at lower energy is expected.

\section*{Acknowledgement}
\vspace{-0.2cm}
SC acknowledges UNESCO and the L'Or\'eal Foundation for the support received through the fellowship \textquoteleft\textquoteleft L'Or\'eal Italia Per le Donne e la Scienza\textquoteright\textquoteright. GM acknowledges the support received through The Grants ASI/INAF n. 2017-14-H.O and SKA-CTA-INAF 2016. SG acknowledges support from Agence Nationale de la Recherche (grant ANR- 17-CE31-0014) and from the Observatory of Paris (Action F\'ed\'eratrice CTA). 

\bibliographystyle{mnras}
\bibliography{References}


\appendix

\section{Theoretical estimate of $\delta$}	
\label{sec:appA}
The temporal dependence of the maximum momentum as described in Eq.~(\ref{eq:pmax0}) can be estimated from a simple theoretical argument which is often used to estimate the maximum energy in the test-particle DSA, namely by equating the acceleration time with the age of the remnant $T_{\rm SNR} = t_{\rm acc}(p)$. Using $t_{\rm acc}(p)=D(p)/u^2_{\rm sh}$ for the acceleration time and writing the diffusion coefficient in terms of the magnetic turbulence, $D= D_B \mathcal{F}^{-1}$, where $D_B= p c/(3B_0)$ is the Bohm diffusion coefficient ($B_0$ being the regular background magnetic field) and $\mathcal{F}$ is the turbulent magnetic energy density per unit logarithmic bandwidth of waves (normalized to the background magnetic energy density), we can write:
\begin{eqnarray} \label{eq:pmaxB}
 p_{\max,0}(t)  \propto \mathcal{F}(t) \, u_{\rm sh}^2(t) \, t \, .
\end{eqnarray}
If there is no magnetic field amplification, the diffusion is determined by the pre-existing magnetic turbulence which is stationary. As a consequence the time dependence is only determined by the shock speed which evolves $\propto t^{-3/5}$ in the ST phase, resulting in $p_{\max,0}(t) \propto t^{-1/5}$. Such a result represents a minimum value for $\delta$, that applies when neither amplification nor damping of magnetic turbulence are taking place. Conversely, if the turbulence is amplified, a steeper time dependence is expected. In the case of resonant streaming instability, for instance, $\mathcal{F} \propto P_{\rm CR} \propto u_{\rm sh}^2$, hence $p_{\max,0}(t) \propto t^{-7/5}$. A similar result holds even in the case of non-resonant instability, which, according to \cite{Bell+2013}, gives $\mathcal{F} \propto u_{\rm sh}^2$. Note, however, that in previous works \citep{Bell2004} it is argued that tension in the field lines limits amplification when $\nabla \times {\bf B} \sim \mu_0 \, {\bf j}_{\rm CR}$, which results in a saturated turbulence with $\mathcal{F} \propto u_{\rm sh}^3$, leading to $\delta=2$ from Eq.~(\ref{eq:pmaxB}).
In addition, if any magnetic damping mechanism is effective in the shock region, like MHD cascade or ion-neutral friction, an even larger value of  $\delta$ is foreseen.

\section{Analytical solution of diffusive transport equation without advection}	
\label{sec:appB}
If the diffusion coefficient is constant in space, Eq.~(\ref{eq:transport_out}) can be reduced to a one-dimensional cartesian problem for the function $g= r \, f_{\rm esc} $, namely
\begin{equation} 
\label{eq:transport_out2}
 \frac{\partial{g(t,r,p)}}{\partial{t}} =   D(p) \frac{\partial^2 g(t,r,p)}{\partial r^2}  \,,
\end{equation}
with the boundary condition $g(t, r=0, p) = 0$ and the initial condition $g(t=0, r, p) = r \, f_{\rm conf}(t_{\rm esc}(p), r, p) \equiv r f_{\rm conf,0}(r, p)$. 
In the following we drop the dependence on $p$.
Now we can use the Laplace transform, $\mathcal{G}(s,r) = \int_{0}^{\infty} e^{-st} g(t,r) dt$, to rewrite Eq.~(\ref{eq:transport_out2}) as an ordinary second order differential equation, i.e.
\begin{equation} 
\label{eq:transport_out3}
  \frac{\partial^2 \mathcal{G}(s,r)}{\partial r^2}  = \frac{s}{D}  \mathcal{G}(s,r) - \frac{r f_{\rm conf, 0}(r)}{D} \,,
\end{equation}
with the boundary conditions $\mathcal{G}(s,0) = \mathcal{G}(s,\infty) = 0$. Because of the boundary conditions, the solution of the associate homogeneous equation is identically zero while the particular solution can be found using standard techniques, by solving:
\begin{equation} 
\label{eq:sol1}
\mathcal{G}(s,r) = e^{-\omega r} \int_{0}^{r} dr' e^{2\omega r'} \int_{r'}^{\infty} dr'' \frac{r'' f_{\rm conf,0}(r'')}{D} e^{-\omega r''}  \, ,
\end{equation}
where $\omega= \sqrt{s/D}$. The radial dependence of the confined density function is provided in Eq.~\eqref{eq:f_in(t,r)}, as it is enclosed in $t'(t,r)$. Three different situations have been explored in \S~\ref{sec:f_esc}, namely: \\

\noindent
{\it i}) $\alpha=4, \sigma=4 \implies f_{\rm conf,0}(r)=\textrm{const}$ (see Eq.~\refeq{eq:f_in(t,r)}): \\
\begin{equation} 
\mathcal{G}(s,r) = \frac{f_{\rm conf,0}}{s} \left[ M e^{(M-r) \omega} - \frac{1+\omega R}{2\omega} \left( e^{(2M-R-r) \omega} - e^{-(R+r) \omega} \right) \right] \, ,
\end{equation}
where $R \equiv R_{\rm esc}(p)$ and $M=\min(r,R)$. Performing the inverse Laplace transform of the latter expression, we finally get
\begin{eqnarray} 
\label{eq:sol2a}
  f_{\rm esc}(t,r) &=& g(t,r) \, r^{-1} =  \hspace{5.cm}   \\ \nonumber
    &=& \frac{f_{\rm conf, 0}}{r} 
    	    \left\{ M \left( {\rm Erfc} \left[ \frac{r-M}{R_d}\right] -1 \right)  \right. + \\ \nonumber
    &+& \frac{R_d}{2 \sqrt{\pi}} 
  		\left( e^{-\left(\frac{r+R}{R_d} \right)^2} -e^{-\left(\frac{r+R-2M}{R_d} \right)^2} \right) + \\ \nonumber 
    &+& \frac{r+R}{2} {\rm Erf} \left[ \frac{r+R}{R_d}\right] - \frac{(r+R-2 M)}{2} {\rm Erf} \left[ \frac{r+R-2M}{R_d}\right]  +\\ \nonumber 	      
    &+& \left. \frac{R}{2} {\rm Erfc} \left[ \frac{r+R}{R_d}\right]
          -  \frac{R}{2} {\rm Erfc} \left[ \frac{r+R-2M}{R_d}\right] 
  \right\} \, .
\end{eqnarray}
With a little algebra, the above solution can be simplified giving the expression in Eq.~(\ref{eq:sol2}), valid for both $r < R$ and $r > R$. \\

\noindent
{\it ii}) $\alpha=4+1/3, \sigma=4 \implies f_{\rm conf,0}(r) \propto r$ (see Eq.~\refeq{eq:f_in(t,r)}): \\
\begin{equation} 
\begin{split}
\frac{\mathcal{G}(s,r)}{k(t_\textrm{esc})} & = \frac{1}{\omega^3 D} e^{-\omega r} \left[ -\frac{2}{\omega} + e^{\omega M} \left( \frac{2}{\omega} + \omega M^2 \right) \right. + \\
& \, \left. + e^{2 \omega M} \left( -\frac{1}{\omega} - \frac{\omega}{2} R^2 - R \right) \right] \, ,
\end{split}
\end{equation}
and the inverse Laplace transform yields
\begin{equation}
\label{eq:unl}
\begin{split}
\frac{rf_{\rm esc}(t,r)}{k(t_\textrm{esc})} & = 2M^2-2Mr+\frac{R_d}{\sqrt{\pi}} \left[ re^{-\frac{r^2}{R_d^2}} + (M-r)e^{-\frac{(M-r)^2}{R_d^2}} \right] + \\
& + \frac{R_d}{\sqrt{\pi}} (r-2M+R) e^{-\frac{(r-2M+R)^2}{R^2_d}} \left[\frac{1}{2} - \frac{R}{\lvert r-2M+R \rvert} \right] + \\
& +  (r^2+\frac{R_d^2}{2}) \textrm{Erf}\left[ \frac{r}{R_d} \right] +  \frac{R_d}{2\sqrt{\pi}} (R-r) e^{-\frac{(r+R)^2}{R^2_d}}  + \\
& +  (2M^2-2Mr+r^2+\frac{R_d^2}{2}) \textrm{Erf}\left[ \frac{M-r}{R_d} \right] + \\
& + \frac{1}{2} (r^2+\frac{R_d^2}{2}) \textrm{Erfc}\left[ \frac{r+R}{R_d} \right] - \frac{1}{2 (r-2M+R)} \times  \\
& \times  \left( 4M^2 + r^2 +2rR +R^2 -4M(r+R) + \frac{R_d^2}{2} \right) \times \\
& \times \left( \lvert 2M-r-R \rvert - (2M-r-R)\textrm{Erf} \left[\frac{2M-r-R}{R_d} \right]  \right) + \\
& +R\textrm{Erfc} \left[\frac{\lvert r-2M+R \rvert}{R_d}  \right] \left( 1 - \frac{R}{2} \frac{1}{\lvert r-2M+R \rvert} \right) \times \\
& \times (r-2M+R) \, , \\
\end{split}
\end{equation}
which can be also formulated as in Eq.~\eqref{eq:fesc43}. \\

\noindent
{\it iii}) for the precursor $f_{\rm conf,0}(r) \propto \delta(r-R_{\rm sh})$ (see Eq.~\refeq{eq:precDelta}): \\
\begin{equation}
\mathcal{G}(s,r) = f_0(p,t_\textrm{esc})  \frac{D_p(p)}{D(p)} \frac{R}{u_{\rm sh}(t_\textrm{esc})}  \frac{1}{2\omega} e^{-\omega (r+R)}(e^{2\omega M}-1) \, ,
\end{equation}
and finally its inverse Laplace transform reads as
\begin{equation}
f_{\rm esc}(t,r) = \frac{f_0(p,t_\textrm{esc})}{\sqrt{\pi}} \frac{R}{R_d} \frac{D_p(p)}{u_{\rm sh}(t_\textrm{esc})r} \left[ \exp^{-( \frac{r+R-2M}{R_d})^2} - \exp^{-( \frac{r+R}{R_d})^2} \right] \, ,
\end{equation}
which is identical to the expression reported in Eq.~\eqref{eq:precFinal}. \\


\section{Self-generated turbulence} 
\label{sec:turbulence}

At this point it is worth discussing in more details the value of the diffusion coefficient expected to be operating outside of the sources. The assumption that the diffusion coefficient in the region around an SNR should be the same as the average Galactic one, as derived from direct measurement of secondary/primary CR ratios \citep{maurin2014}, does not have a strong justification. Indeed, the latter one mainly measures the diffusion as it occurs in the Galactic magnetic halo whose transport properties can be remarkably different from the regions around SNRs. On the other hand, it is easy to imagine mechanisms able to enhance the magnetic turbulence around an SNR, especially when it originates from a core-collapse (CC) explosion. First of all, the circumstellar environment can be modified by the pressurized bubble produced by the progenitor wind. In addition, many CC-SNRs explode in OB associations, where frequent SN explosions, as well as winds from massive stars, can easily enhance the local magnetic turbulence in a region of tens of parsecs, resulting in a suppressed diffusion coefficient. 

Beyond those mechanisms, also instabilities produced by run-away CRs can amplify the magnetic turbulence, suppressing the diffusion coefficient by orders of magnitudes. In particular, the role of resonant instability produced by escaping particles has been studied by several authors \citep{Ptuskin-Zirakashvili-Plesser:2008, Yan-Lazarian-Schlickeiser:2012, Malkov+2013, Evoli2018, Nava+2019}, showing that a suppression by one to two orders of magnitude is possibly achieved in the energy range below $\sim 1$ TeV, inside a region up to tens of parsecs from the SNR. Nevertheless, in the case that a large fraction of neutral Hydrogen is populating the CSM, the amplification effect can be reduced by the ion-neutral friction \citep{Kulsrud-Pearce:1969} resulting in a much smaller level of turbulence as shown by \cite{Nava+2016} and \cite{DAngelo+2018}. A close comparison between those results and our findings is not obvious, mainly because, their recipes for particle escape is different from ours. Moreover, our model is spherically symmetric, while they both assume a diffusion along a one dimensional flux tube.
Nevertheless, we can use the spatial CR gradient obtained with a given assumption on $D_{\rm out}$ to estimate {\it a posteriori} the level of self-generated turbulence due to streaming instability. Such a procedure is very similar to the one already used by \cite{Yan-Lazarian-Schlickeiser:2012}.
Even if such a calculation is not a self-consistent one, it can show whether or not the streaming instability can be responsible for the reduction of $D_{\rm out}$. It is worth stressing that one should account for the duration of the wave amplification process: on a general ground, one can expect that a suppression of the diffusion coefficient with respect to the average Galactic value is achieved within few escape times, but later on, when the CR density diminishes, also the amplification of the magnetic turbulence fades. In order to facilitate the comparison among the remnant age and the escape time of particles at different energy, we report in Tab.~\ref{tab:tab2} the expected escape time, computed according to Eq.~\eqref{eq:tesc} and benchmark values reported in Tab.~\ref{tab:tab1}.

\begin{table}
\centering
\footnotesize
\caption{Escape times for particles of different momentum from a SNR evolving according to the benchmark values in Tab.~\ref{tab:tab1}. The parametrization of escape time adopted here follows Eq.~\eqref{eq:tesc}, with $\delta=3$.}
\label{tab:tab2}
\begin{tabular}{cc}
\hline	
$p$ (GeV/c)  &  $t_{\rm esc}$ (yr) \\
\hline
$10$ & $7.4 \times 10^4$ \\
$10^2$ & $3.4 \times 10^4$ \\
$10^3$ & $1.6 \times 10^4$ \\
$10^4$ & $7.4 \times 10^3$ \\
$10^5$ & $3.4 \times 10^3$ \\
\hline  
\end{tabular}
\end{table}

In order to estimate the level of self-generated turbulence, we need to compare the amplification rate by resonant streaming instability with the damping rate of Alfv\'en waves. The amplification rate of waves with wavenumber $k$ in resonance with particles of Larmor radius $r_L$ as due to streaming instability is~\citep{Skilling1971}
\begin{equation}
\label{eq:GammaCR}
\Gamma_\textrm{CR} (k) = \frac{16 \pi^2}{3} \frac{v_A}{B_0^2 \mathcal{F}(k)} \left[ p^4 v(p) \frac{\partial f}{\partial r} \right]_{p=p_\textrm{res}} \, ,
\end{equation} 
where $B_0$ is the intensity of the background magnetic field and $v_A=B_0/\sqrt{4\pi n_i m_i}$ is the Alfv\'en speed ($m_i$ and $n_i$ being respectively the mass and density of the ions in the CSM). Here, $\mathcal{F}(k)$ is the normalized energy density of magnetic turbulence per unit logarithmic wavenumber $k$, calculated at the resonant wavenumber $k_\textrm{res}= 1/r_L(p_\textrm{res})$. An useful way to write $\mathcal{F}(k)$ is by using the Bohm diffusion coefficient, $\mathcal{F}(k)=D_B/\hat{D}$, where $\hat{D}$ is the self-generated diffusion coefficient. \\

\begin{figure}
\centering
\includegraphics[width=0.48\textwidth]{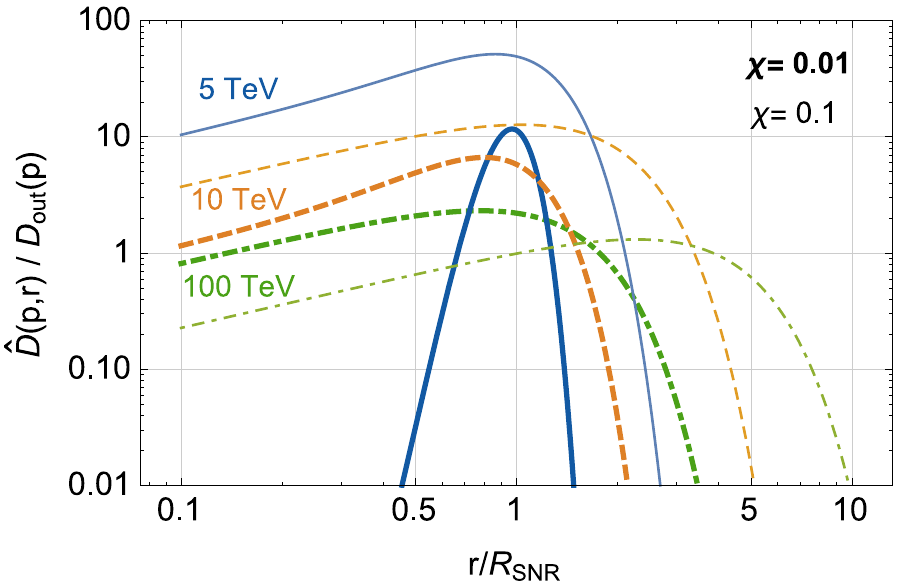}
\includegraphics[width=0.48\textwidth]{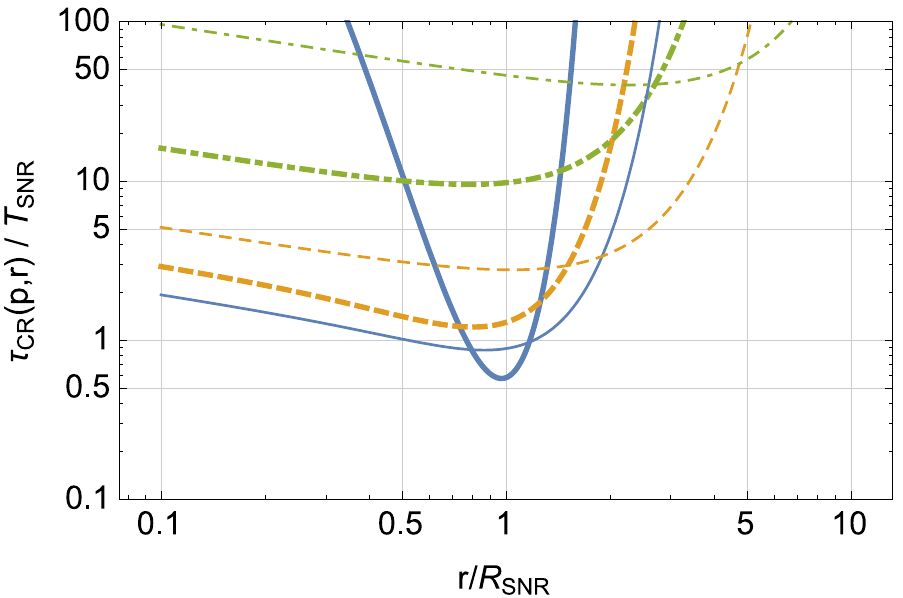}
\caption{\textit{Top:} spatial dependence of self-generated diffusion coefficient $\hat{D} (p,r)$ divided by $D_{\rm out}(p)$ for $\chi = 0.1$ (thin lines) and $\chi = 0.01$ (thick lines), calculated by setting an acceleration spectrum with slope $\alpha=4$ for the benchmark parameter values reported in Tab.~\ref{tab:tab1}. The parametrization of escape time adopted here follows Eq.~\eqref{eq:tesc}, with $\delta=3$. The three sets of lines correspond to three different particle energies: $5$~TeV (solid), $10$~TeV (dashed) and $100$~TeV (dot-dashed). \textit{Bottom:} corresponding excitation time for the streaming instability in unit of SNR age ($10^4$ yr) and for the same energy values as the left panel.}
\label{fig:Dself} 
\end{figure}

Concerning the damping mechanisms in a completely ionized plasma, several processes might affect the propagation of magnetic waves, as turbulent cascading, wave-particle interactions (e.g. non-linear Landau damping \citep{kulsrud1978}) and wave-wave interactions (e.g. the interaction among self-generated waves and background turbulent perturbations \citep{farmer2004,lazarian2016}). For the sake of simplicity, we will limit the following analysis to the cascade damping, namely the Kolmogorov-type energy cascade towards large wavenumbers. As a consequence, the resulting turbulence should be considered as a rough estimate of that actually developing in the plasma. Within the cascade process, the damping of Alfv\'enic waves occurs non-linearly (NLD) at a rate~\citep{Ptuskin-Zirakashvili:2003}
\begin{equation} 
\label{eq:Gamma_damp}
  \Gamma_{\rm NLD}(k) = (2 c_k)^{-3/2} \, k v_A  \, \sqrt{\mathcal{F}(k)} \, ,
\end{equation}
where $c_k= 3.6$ is called Kolmogorov constant. Now, by equating $\Gamma_{\rm CR}$ with $\Gamma_{\rm NLD}$ one gets
\begin{equation} 
\label{eq:F}
  \mathcal{F}(k) = \frac{D_B}{\hat{D}} 
          = 2c_k \left[ \frac{16}{3} \frac{\pi^2}{B_0^2} \left(p^4 v(p) \frac{\partial f_{\rm esc}}{\partial r} \right)_{p=p_\textrm{res}} r_L \right]^{2/3}  \,.
\end{equation} 
Assuming the same benchmark values for the parameters as in Tab.~\ref{tab:tab1} with a background magnetic field $B_0 \simeq 3 \, \mu$G and $\alpha=4$, we calculated the ratio $D_{\rm out}/\hat{D}$ for $\chi =0.1$ and $\chi =0.01$. Results are shown in the top panel of Fig.~\ref{fig:Dself}. As visible, in both cases, the level of self-generated turbulence is such that $\hat{D} \lesssim D_{\rm out}$ for $p c  \lesssim 100$ TeV in a region of few times the size of the SNR. On the other hand, the timescale to excite the instability, reported in the bottom panel of the same Figure, is smaller, or comparable, to the SNR age only for energy lower than $\sim 10$ TeV. As a consequence, below such energy the resonant streaming instability is able to reduce the diffusion coefficient, but only in a spatial region close to the SNR radius.

\end{document}